\documentclass[aps,preprint,showpacs,groupedaddress,floatfix]{revtex4}

\usepackage{graphicx}
\usepackage{dcolumn}
\usepackage{bm}
\usepackage{CJK}

\begin{document}
\begin{CJK*}{GBK}{song}

\title{Gamow-Teller response and its spreading mechanism in doubly magic nuclei}
\author{Y. F. Niu   $^{1,3}$}
\email{nyfster@gmail.com}
\author{G. Col\`{o} $^{2,1}$}
\email{gianluca.colo@mi.infn.it}
\author{E. Vigezzi $^{1}$}
\email{vigezzi@mi.infn.it}

\affiliation{$^1$ INFN, Sezione di Milano, via Celoria 16,
I-20133 Milano, Italy}
\affiliation{$^2$ Dipartimento di Fisica, Unversit\`{a} degli Studi
di Milano, via Celoria 16, I-20133 Milano, Italy}
\affiliation{$^{3}$ Institute of Fluid Physics, China Academy of
Engineering Physics, Mianyang 621900, China}

\date{\today}

\begin{abstract}
The scope of the paper is to apply a state-of-the-art beyond mean-field model to the
description of the Gamow-Teller response in atomic nuclei. This topic  recently attracted
considerable renewed interest,  due, in particular, to the possibility of performing experiments in
unstable nuclei. We study the cases of
$^{48}$Ca, $^{78}$Ni, $^{132}$Sn and $^{208}$Pb.
Our model is based on a fully self-consistent Skyrme Hartree-Fock plus random phase
approximation. The same Skyrme interaction  is used  to  calculate the coupling between particles and vibrations,
which leads to the mixing of  the Gamow-Teller resonance with a set of doorway states
and to its fragmentation.
We compare our results with available experimental data.
The microscopic coupling mechanism is also discussed in some detail.
\end{abstract}
\pacs{
 21.60.Jz, % Nuclear Density Functional Theory and extensions (includes Hartree-Fock and random-phase approximations)
 23.40.Hc, % Relation with nuclear matrix elements and nuclear structure (under item 23.40.-s ¦Â decay; double ¦Âdecay; electron and muon capture  )
 24.30.Cz, % Giant resonances
 25.40.Kv  % Charge-exchange reactions
 } \maketitle
\date{today}
%%%%%%%%%%%%%%%%%%%%%%%%%%%%%%%%%%%%%%%

\section{Introduction}

Spin-isospin resonances are not a new subject, yet they still capture the interest of researchers,
both  experimentalists and  theorists, not only in nuclear physics but also in particle physics and astrophysics. There exist some review papers \cite{Osterfeld1992, Ichimura2006}, that do not cover, however, the most recent developments. From the experimental point of view, the forefront of such kind of research is the exploration of the spin-isospin modes in exotic nuclei (mainly neutron-rich and possibly drip-line isotopes). This is nowadays possible due to the advent of radioactive beam facilities, and a dedicated effort focused on charge-exchange reactions leading to spin-isospin modes is ongoing, e.g., in Japan and in the USA. The Gamow-Teller (GT) excitations are the main goal but
spin-dipole
or other multipoles are also of interest.

Effective nucleon-nucleon  forces are poorly constrained in
the spin-isospin channel, and they often display unphysical
ferromagnetic instabilities in nuclear matter at high densities
that need to be cured (cf. \cite{Navarro2013,Chamel2010}
and references therein).
A systematic exploration of
spin-isospin transitions,
from light to medium-heavy systems and from the neutron-deficient
to the neutron-rich side, is
needed, in order to tune such
effective nucleon-nucleon interactions in the nuclear medium and to study the nuclear equation of state.
It has been shown, for instance, that the spin-dipole strength is a good indicator of the neutron skin (cf. \cite{Colo2014} and references therein). Most of the GT strength  in the $t_-$ channel lies at high excitation energy in stable nuclei; however, as the neutron excess increases, one expects that a sizable part of it can move down into the $\beta$-decay window \cite{Sagawa1993}. To what extent this happens,
and/or brings new information on the spin-isospin part of the nuclear Hamiltonian, is still an open question: although it  can be at present mainly tackled in light nuclei such as $^8$He \cite{Kobayashi:2014} and $^{12}$Be,
 there are attempts towards the neutron-rich side in heavy systems as well (for instance, in the
case of $^{132}$Sn).

Spin-isospin excitations play an important role in the weak-interaction
processes such as electron capture, $\beta$ decay, and neutrino-nucleus
reactions. Therefore, the knowledge of selected  transitions
in a specific series of nuclei, or in specific
regions of the nuclear chart, is of great interest for nuclear astrophysics \cite{Langanke2003,Janka2007}. A clear and well-known example is that of core-collapse supernovae. In this case, the electron capture rates govern the evolution
of the system and consequently, GT transition matrix elements must be accurately known in the iron region  \cite{Langanke2003,Janka2007,Paar2009,Niu2011,Fantina2012}. The $\beta$-decay half-lives set the time scale of the rapid neutron capture process ($r$ process), and hence influence the production of heavy elements in the universe \cite{Burbidge1957,Qian2007,Niu2013}. Last but not least, a very accurate knowledge of spin-isospin matrix elements is also instrumental to extract the properties of the neutrinos from the measured half-life of double-$\beta$ decay \cite{Avignone2008,Vergados2012}.

From the theoretical point of view, in the last two decades there has been significant progress
in microscopic models aimed at the description of collective excitations such as the charge-exchange GT
and spin-dipole resonances. Most nuclei in the nuclear chart can be studied by  models based
on self-consistent mean-field or density functional theory. At present,
they can be employed in the form of a Hartree [or Hartree-Fock (HF)] approximation for the ground-state
plus charge-exchange random phase approximation (RPA) to determine the main resonance properties.
This can be done using
Skyrme \cite{Auerbach:1984,Hamamoto-Sagawa:2000,Bender2002,Fracasso2007,
Bai2009,Roca-Maza2012} or Gogny \cite{Martini:2014} effective Hamiltonians,
as well as using covariant effective Lagrangians \cite{DeConti:2000,Paar2004,Liang2008}. Some of these Hamiltonians or Lagrangians can reproduce the experimentally observed mean energy, and the fraction of the exhausted sum rule, for the GT and spin-dipole resonances, although they are  not based on the same physical picture.

Self-consistent mean-field models have also limitations. It is well known that they cannot account
for the  spreading widths of giant resonances.
The problem of the fragmentation of the GT strength has been addressed
using second RPA \cite{Drozdz:1990} and the quasiparticle-phonon model
\cite{Kuzmin1984} (see also Ref. \cite{Dang:1997}).
Generally speaking,  models based on  particle-vibration coupling (PVC)
are quite effective in reproducing the giant resonance widths
\cite{Bertsch1983}.
We apply such an approach in the present  paper, which is a follow-up of Ref. \cite{Niu2012}.
As described in that work, some of us implemented a scheme based on the fully self-consistent Skyrme HF plus RPA,
in which specific diagrams associated with PVC corrections at lowest order are introduced,
and no further approximation has been done. In particular, the same Skyrme force is used to calculate  the single-particle levels and
the RPA spectrum (including both  the GT state and the low-lying surface vibrations to be coupled to it),
as well as the PVC vertices (see next section). While the model is similar to the one of Ref. \cite{Colo1994}, it includes many improvements; in particular, all the terms of the Skyrme interaction are taken into account.
 Our goal here is to see if we can reproduce the line shape of the strength function,
and the associated spreading width, observed in  different nuclei and  in different mass regions. In particular, we wish to make predictions for exotic nuclei  for which experiments have already  been carried out and not yet analyzed, or are planned. We share part of our motivations with the recent work of Ref. \cite{Litvinova2014} which adopts
a model similar to ours, but based on a covariant description. We also present  a detailed discussion of the spreading mechanism within our approach.
The damping width of giant resonances has also a contribution coming
from the escape width; that is, from the nucleon emission. In the GTR
case, experiment indicates that the escape width is very small
(of the order of $\approx$4\% of the total width in $^{208}$Pb, cf. Table 8.3 of
\cite{Harakeh-vanderWoude:2001}). Consequently, we have not
included the continuum coupling in the present paper, and we compare directly the calculated spreading
width with the total experimental width.

The outline of our work is as follows. In Sec. \ref{formalism}   a short account of the formalism is presented.
In Sec. \ref{discussion1} the GT strength distributions and cumulative sums calculated for nuclei $^{208}$Pb, and $^{48}$Ca are compared with the experimental data, while predictions are provided for  $^{132}$Sn and $^{78}$Ni.
%\textcolor{blue}
{The overall features of the GT resonance (GTR) and phonons are summarized in Sec. \ref{discussion1.5}.} The  PVC mechanism for the spreading and fragmentation of GTR is discussed in Sec. \ref{discussion2}. Finally, the main conclusions of this work are summarized in Sec. \ref{conclu}.

\section{Formalism}\label{formalism}

We employ the same formalism as in Ref. \cite{Niu2012}, and here we only recall its essential points. We first carry out a self-consistent HF+RPA calculation of the GT strength, using a standard Skyrme interaction.
% $V$.
The HF equations are solved in coordinate space on a radial mesh of size $0.1$ fm, within a spherical
box having a radius equal to $21$ fm. The continuum is discretized by requiring vanishing boundary conditions for the wave functions at the edge of this box. A set of RPA eigenstates $|n\rangle$ for the GT excitations are obtained by the diagonalization of the RPA matrix.
% \textcolor{blue}
{Forward-going and backward-going amplitudes associated with the RPA eigenstates $|n\rangle$ will be denoted by $X^{(n)}_{ph}$ and $Y^{(n)}_{ph}$, respectively.} Single-particle states up to 100 MeV have been included in the particle-hole (p-h) configuration space.  Within PVC, the RPA strength will be shifted and redistributed through the coupling to a set of doorway states, denoted by $|N\rangle$, made of a p-h excitation coupled to a collective vibration of angular momentum $L$. The properties of these collective vibrations, i.e., phonons $|nL\rangle$ are, in turn, obtained by computing the RPA response with the same Skyrme interaction, for states of natural parity $L^{\pi} = 0^+$, $1^-$, $2^+$, $3^-$, $4^+$, $5^-$, and $6^+$. For the PVC model space, we have retained the phonons with energy less than 20 MeV and absorbing a fraction of the total isoscalar or isovector strength larger than $5\%$, and included intermediate particle states up to an energy of 100 MeV.

\begin{figure}
\centerline{
\includegraphics[scale=0.65,angle=0]{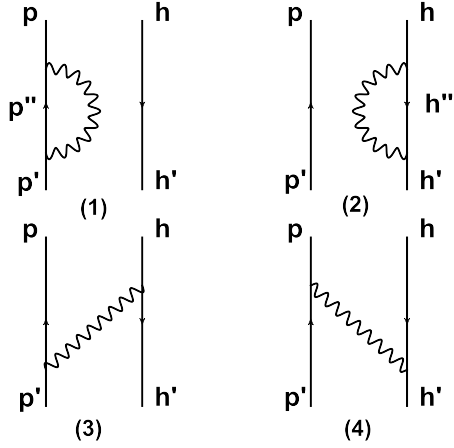}
} \caption{Diagrammatic representation of the four terms whose sum
gives the matrix element $W^\downarrow _{ph,p'h'}$. The analytic
expressions are shown in Eq. (\ref{Wdown1}).} \label{fig0}
\end{figure}

The GT strength associated with RPA+PVC, is given by
\begin{equation}
\label{strength}
  S(\omega)
  = -\frac{1}{\pi} {\rm Im} \sum_\nu \langle 0 | \hat O_{\rm GT^\pm} | \nu \rangle
  ^2 \frac{1}{\omega - \Omega_\nu + i(\frac{\Gamma_\nu}{2}+\Delta)},
\end{equation}
where the GT operator is $\hat O_{\rm GT ^{\pm}} = \sum_{i=1}^{A} \mathbf{\sigma}(i) \tau_{\pm}(i)$. In our calculation, we will only focus on the GT$^-$ excitations. $|\nu\rangle$ denote the eigenstates [associated with the complex eigenvalues
$\Omega_\nu - i\frac{\Gamma_\nu}{2}$ and eigenvectors $(F^{(\nu)}, \bar{F}^{(\nu)})$] that
are obtained by diagonalizing the energy-dependent complex matrix

\begin{equation}
  \label{RPAmatrix}
    \left( \begin{array}{cc} {\cal D} + {\cal A}_1(\omega) & {\cal
    A}_2(\omega) \\ -{\cal A}_3(\omega) & -{\cal D} - {\cal
    A}_4(\omega) \end{array} \right) \left( \begin{array}{c}
    F^{(\nu)} \\ \bar{F}^{(\nu)} \end{array} \right) = (\Omega_\nu - i
    \frac{\Gamma_\nu}{2}) \left( \begin{array}{c}
    F^{(\nu)} \\ \bar{F}^{(\nu)} \end{array} \right).
\end{equation}
Here, ${\cal D}$ is a diagonal matrix containing the positive RPA eigenvalues, and the ${\cal A}_i$ matrices are associated with the coupling to the doorway states. The expressions of ${\cal A}_i$ in the RPA basis $|n\rangle$ are given by
\begin{eqnarray}
  ({\cal A}_1) _{mn} &=& \sum_{ph,p'h'} W^\downarrow _{ph,p'h'} (\omega) X_{ph}^{(m)} X^{(n)}_{p'h'}
  + W^{\downarrow*} _{ph,p'h'} (-\omega) Y_{ph}^{(m)} Y^{(n)}_{p'h'}, \\
  ({\cal A}_2) _{mn} &=& \sum_{ph,p'h'} W^\downarrow _{ph,p'h'} (\omega) X_{ph}^{(m)} Y^{(n)}_{p'h'}
  + W^{\downarrow*} _{ph,p'h'} (-\omega) Y_{ph}^{(m)} X^{(n)}_{p'h'}, \\
  ({\cal A}_3) _{mn} &=& \sum_{ph,p'h'} W^\downarrow _{ph,p'h'} (\omega) Y_{ph}^{(m)} X^{(n)}_{p'h'}
  + W^{\downarrow*} _{ph,p'h'} (-\omega) X_{ph}^{(m)} Y^{(n)}_{p'h'}, \\
  ({\cal A}_4) _{mn} &=& \sum_{ph,p'h'} W^\downarrow _{ph,p'h'} (\omega) Y_{ph}^{(m)} Y^{(n)}_{p'h'}
  + W^{\downarrow*} _{ph,p'h'} (-\omega) X_{ph}^{(m)} X^{(n)}_{p'h'},
\end{eqnarray}
where $W^{\downarrow}$ reads
\begin{equation}
  W^{\downarrow}_{ph,p'h'} (\omega) = \sum_N \frac{\langle ph |V|N\rangle \langle N | V | p'h' \rangle }{\omega
  -\omega_N}.
\end{equation}
The matrix elements are given by the sum of the four Feynman diagrams represented in Fig. \ref{fig0}, whose analytic expressions are
\begin{eqnarray}
 \label{Wdown1}
 W^{\downarrow }_{1php'h'} &=&\delta_{hh'}\delta_{j_p j_{p'}} \sum_{p'',nL}
  \frac{1} {\omega-(\omega_{nL}+\epsilon_{p''}-\epsilon_h)+i\Delta}
  \frac{  \langle p || V || p'',nL
  \rangle \langle p' || V || p'',nL
  \rangle }{\hat{j}_p^2} ,\nonumber\\
  W^{\downarrow }_{2php'h'} &=& \delta_{ p {p'}}   \delta_{j_h j_{h'}} \sum_{h'',nL}
  \frac{1} {\omega-(\omega_{nL}-\epsilon_{h''}+\epsilon_p)+i\Delta}\frac{  \langle h'' || V || h,nL
  \rangle \langle h'' || V || h',nL
  \rangle }{\hat{j}_h ^2} ,\nonumber\\
  W^{\downarrow }_{3php'h'} &=& \sum_{nL}
  \frac{(-)^{ j_{p}-j_{h'}+J+L}}{\omega-(\omega_{nL}+\epsilon_{p}-\epsilon_{h'})+i\Delta}
  \left\{ \begin{array}{ccc} j_p & j_h & J \\ j_{h'} &
  j_{p'} & L \end{array} \right\}
   \langle p' || V || p, nL  \rangle \langle h' || V || h,nL
   \rangle, \nonumber\\
   W^{\downarrow }_{4php'h'}&=& \sum_{nL}
  \frac{(-)^{ j_{p'}-j_{h}+J+L}}{\omega-(\omega_{nL}+\epsilon_{p'}-\epsilon_{h})+i\Delta}
  \left\{ \begin{array}{ccc} j_p & j_h & J \\ j_{h'} &
  j_{p'} & L \end{array} \right\}
   \langle p || V || p',nL \rangle  \langle h || V || h' , nL
  \rangle.\nonumber\\
\end{eqnarray}
In the above formulas, $p$ and $h$ label particle and hole states, respectively. The corresponding angular momentum and single-particle energies are given respectively by $j_p$, $j_h$ and $\epsilon_p$, $\epsilon_h$. ${\hat{j}_i^2}$ is a shorthand notation for $2j_i+1$, while $\omega_{nL}$ denotes the energy of the phonon state $|nL\rangle$. The averaging parameter $\Delta$ is introduced to avoid singularities in the denominator of Eq. (\ref{Wdown1}), and a convenient practical value is $\Delta= 200$ keV. Such a value is usually smaller than $\Gamma_{\nu}/2$ and does not affect  appreciably the RPA+PVC calculation of the strength in Eq. (\ref{strength}) (it was in fact neglected in the calculations of the strength in Ref. \cite{Niu2012}). In the following, we shall also show calculations with larger values of $\Delta$, in order to simulate the experimental resolution. A larger value of $\Delta$ can also effectively take into account the coupling to more complex configurations not included in the current model.

%---------------------------------------------------------------------------------------------------------
%----------------------------------------------------------------------------------------------------------

A useful approximation to the full diagonalization of the matrix in Eq. (\ref{RPAmatrix})
can be obtained by retaining only the diagonal term $({\cal A}_1) _{mm}$, which is defined as self-energy $\Sigma_m$. If only a single pronounced GTR peak having energy $E_{\rm RPA}$ and strength $|\langle 0 | \hat O_{\rm GT^-} |\rm GTR \rangle|^2$
from the RPA calculation is considered, one obtains the following approximate expression for the strength of Eq. (\ref{strength})
\begin{equation}
 \label{strength2}
  S(\omega)
  = \frac{1}{\pi} \frac{\frac{\Gamma_{\rm GTR}(\omega)}{2}+\Delta}{[\omega-\Omega_{\rm GTR}(\omega)]^2+ (\frac{\Gamma_{\rm GTR}(\omega)}{2}+\Delta^2)} | \langle 0 | \hat O_{\rm GT^-} |{\rm GTR}\rangle|^2,
\end{equation}
where
\begin{equation}
\label{Omega}
\Omega_{\rm GTR}(\omega) =    E_{\rm RPA} +{\rm Re}[({\cal A}_1)_{\rm GTR,GTR}(\omega)]
\end{equation}
and
\begin{equation}
\label{Gamma}
 \Gamma_{\rm GTR}(\omega)= -2{\rm Im}[({\cal A}_1)_{\rm GTR, GTR}(\omega)].
\end{equation}
A simple perturbative expression for the strength function Eq. (\ref{strength2}) can be obtained by putting $\omega = E_{\rm RPA}$ in Eqs. (\ref{Omega}) and (\ref{Gamma}). However, a much better approximation can be obtained by solving Eq. (\ref{Omega}) with $\Omega_{\rm GTR}(\omega) = \omega$ self-consistently. The effectiveness and usefulness of these approximations will be further discussed in Sec. \ref{discussion2}.

%-------------------------------------------------------------------------------

\section{Results and Discussions}

\subsection{Gamow-Teller strength distributions and cumulative sums}
\label{discussion1}

%---------------------------------------------------------------------------------------------------------
\begin{figure*}
\includegraphics[scale=0.3,angle=0]{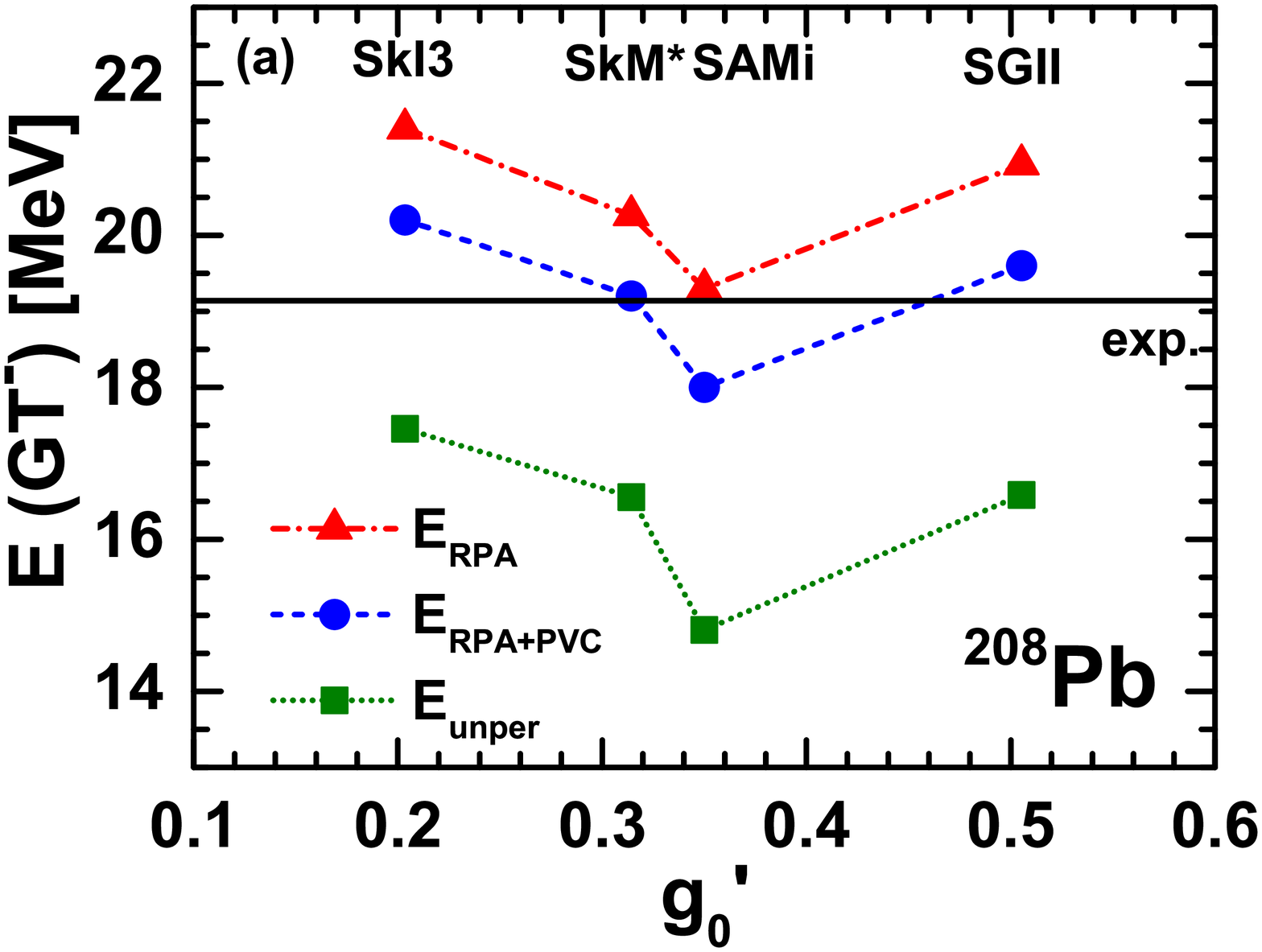}
\includegraphics[scale=0.3,angle=0]{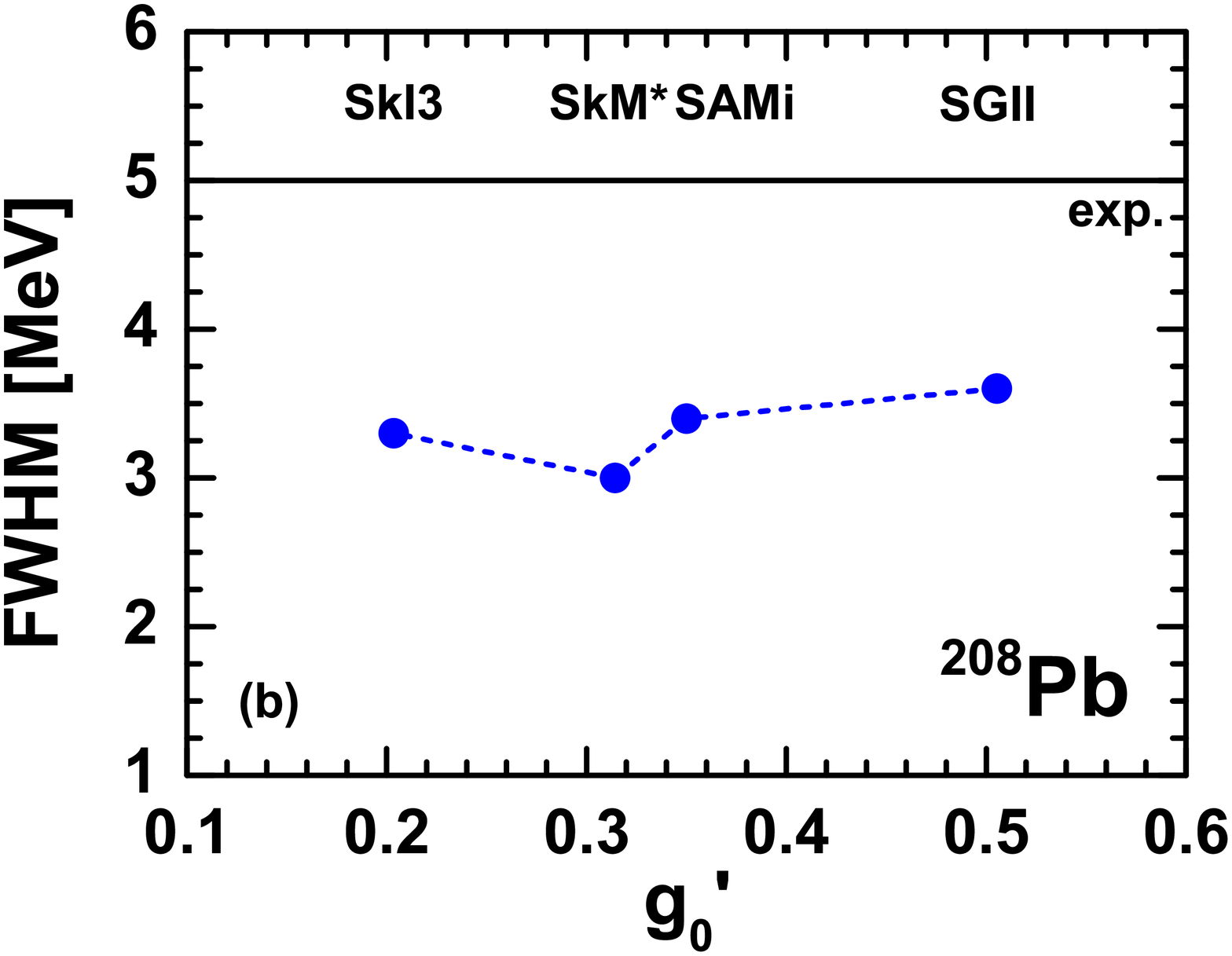}
\caption{(Color online)
Gamow-Teller resonance peak energy [panel (a)]
and FWHM [panel (b)] in $^{208}$Pb, calculated within HF (squares), RPA (triangles), and RPA+PVC (circles) approaches with the Skyrme interactions SkI3, SkM*, SAMi, and SGII.
For convenience, they are shown as a function of the Landau parameter $g_0'$ associated with each force. The unperturbed HF energy $E_{\rm unper}$ is the weighted average of the two main configurations $\nu i_{13/2} \rightarrow \pi i_{11/2}$ and $\nu h_{11/2} \rightarrow \pi h_{9/2}$. The experimental peak energy and width are shown by the black straight lines.} \label{fig01}
\end{figure*}
%----------------------------------------------------------------------------------------------------------

Our calculations can be performed using different Skyrme parameter sets. Consequently,
before discussing in detail the results for the GT strength functions in a series of nuclei,
some considerations about the interaction dependence are in order.
We display in Fig. \ref{fig01} the GTR peak energy calculated in $^{208}$Pb within the
HF, RPA, and RPA+PVC approaches, as well as the GTR full width at half maximum (FWHM) calculated within the RPA+PVC approach, with the Skyrme forces SkI3 \cite{Reinhard1995}, SkM* \cite{Bartel1982}, SAMi \cite{Roca-Maza2012} and SGII \cite{Giai1981}, as a function of the Landau parameter $g_0'$ associated with each force,
and using the value $\Delta=$ 1 MeV for the averaging parameter.
The black straight line in the left panel
denotes the experimental peak energy with respect to the parent nucleus (19.2 MeV),
while the experimental
width (5 MeV) is represented in the same way in the right panel. As is well known,
the unperturbed HF peaks underestimate the peak energy by several MeV. The residual interaction
included in the RPA calculation
raises these values
by 4-5 MeV, overestimating the experimental energy by $\approx$ 1-2 MeV, except for the interaction SAMi~\cite{Roca-Maza2012},
a newly proposed interaction with a good description of the nuclear spin-isospin properties,
that was fitted so as to reproduce the experimental value at RPA level.
The inclusion of PVC acts in a similar way in the four cases. The peak energies are shifted down
by about 1.2 MeV and acquire a width. The FWHM is equal to $\approx$ 3.5 MeV. More precisely, the smallest effect from PVC (1.1 MeV energy shift and 3 MeV FWHM) is obtained in the case of SkM*, and the largest one in the case of
SGII (1.3 MeV energy shift and 3.6 MeV FWHM). We conclude that the effects of particle-vibration coupling are only weakly dependent on the chosen Skyrme set. The best agreement with experiment is obtained with SkM* and SGII, and we will adopt SGII in the rest of our work. The results calculated by using the interaction
SAMi will also be presented for a more detailed comparison in the case of $^{208}$Pb.

%---------------------------------------------------------------------------------------------------------
\begin{figure*}
\includegraphics[scale=0.19,angle=0]{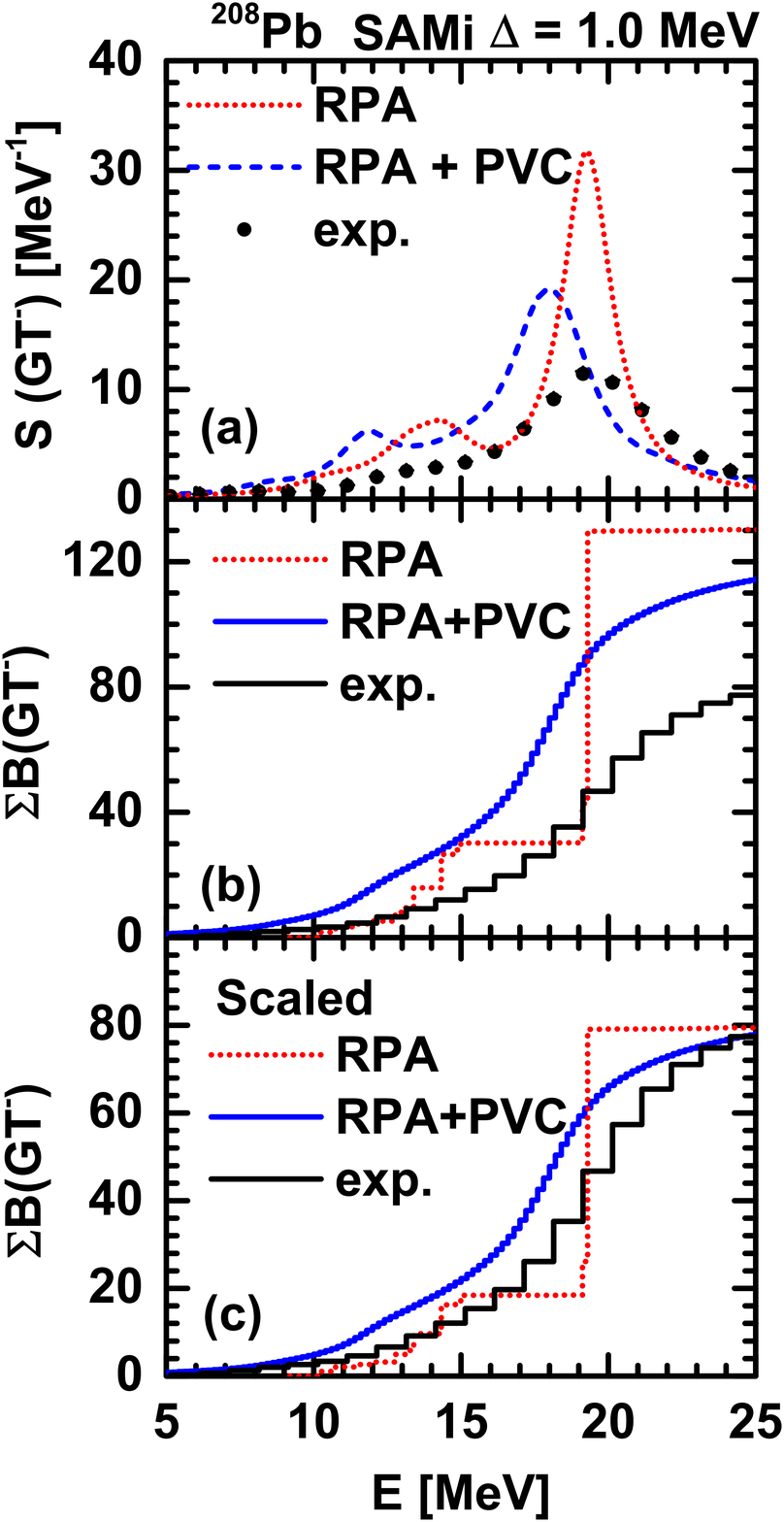}
\includegraphics[scale=0.19,angle=0]{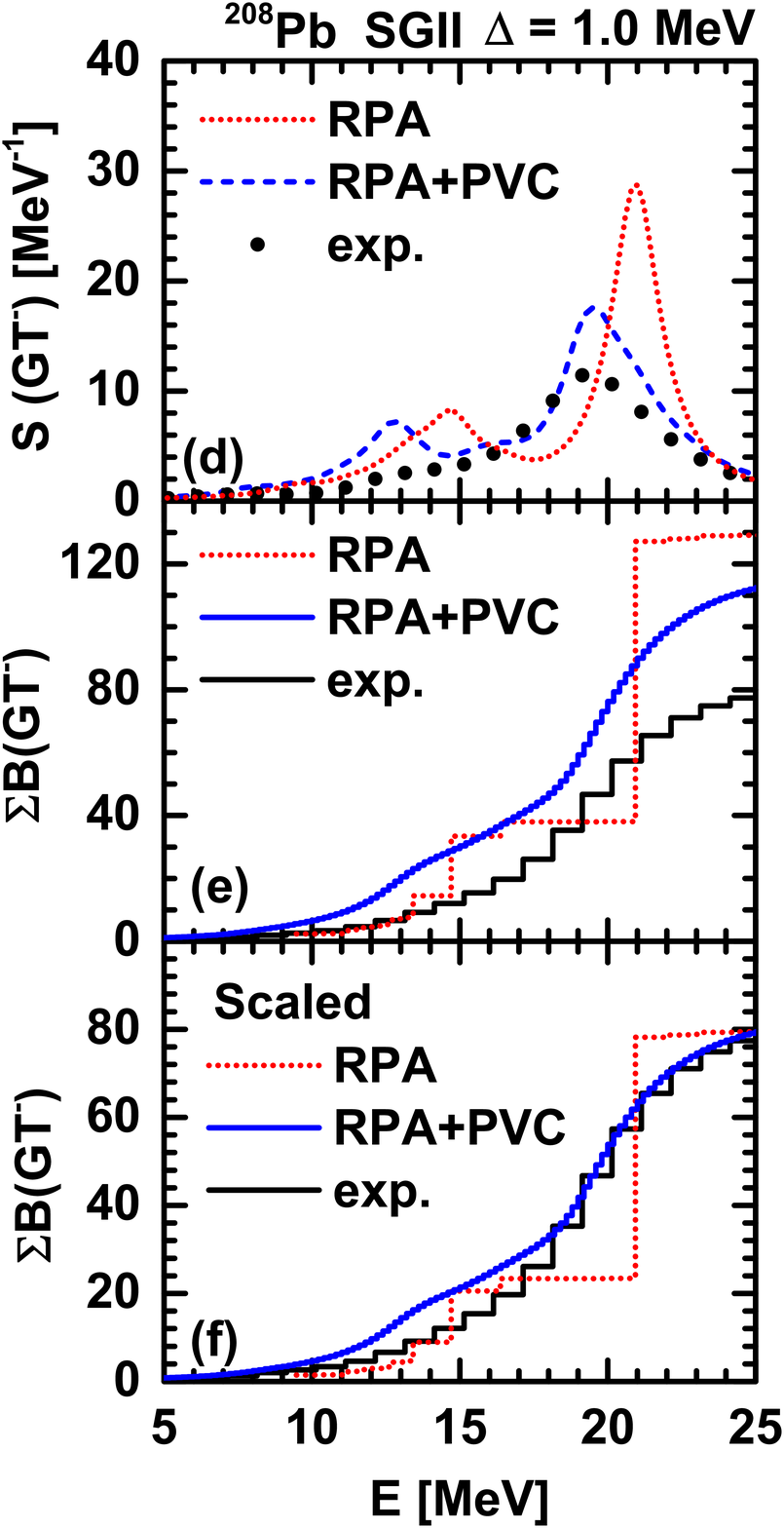}
\includegraphics[scale=0.19,angle=0]{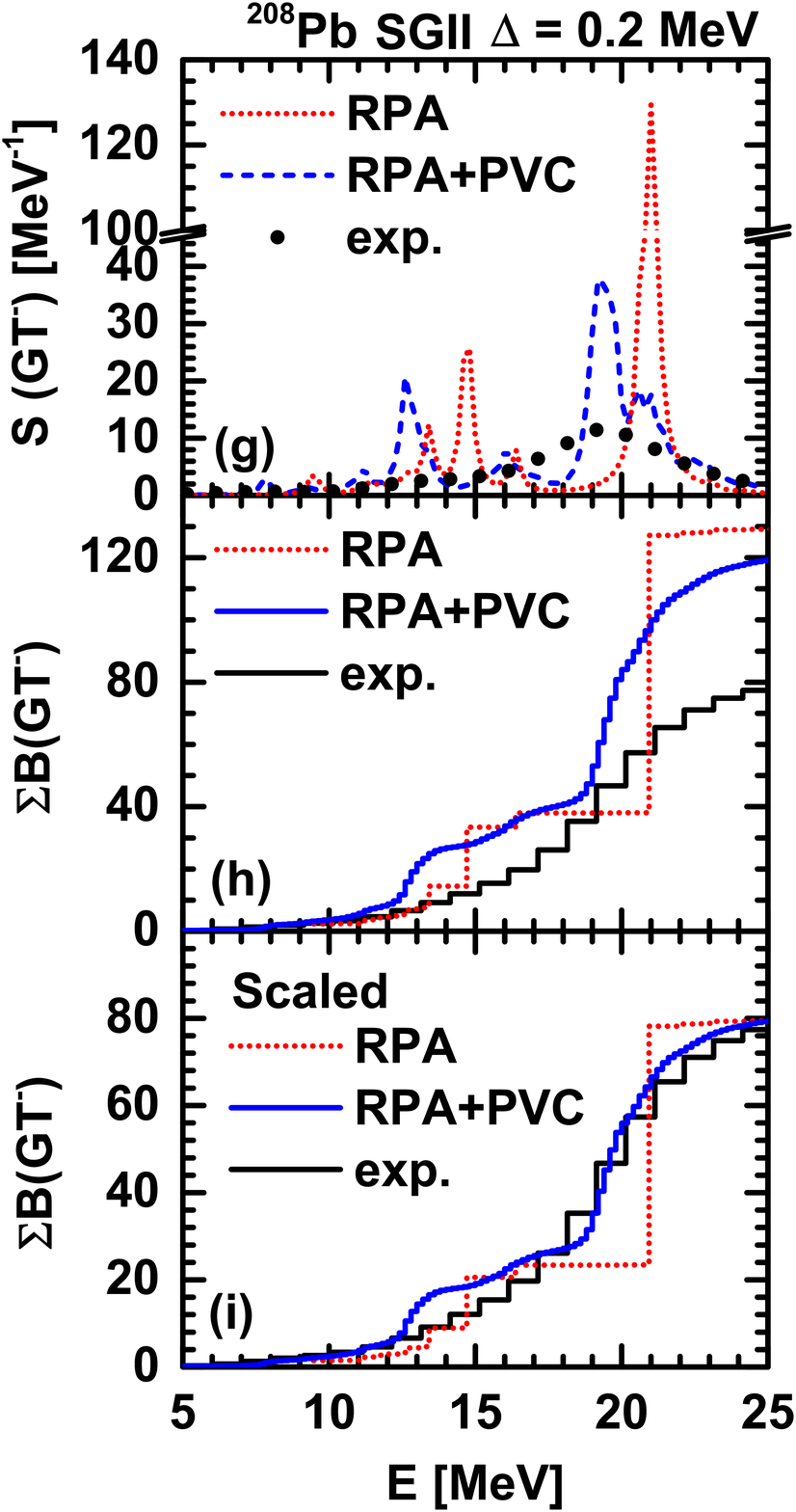}
\caption{(Color online) Gamow-Teller strength distributions [panels (a), (d), and (g)], their cumulative sums
[panels (b), (e), and (h)] and scaled cumulative sums [panels (c), (f), and (i)], calculated with the Skyrme
interactions SAMi (first column) and SGII (second and third columns) for the nucleus $^{208}$Pb. The red-dotted
lines and blue-dashed lines denote the results of RPA and RPA+PVC model, respectively. The smearing parameter
$\Delta$ in the calculations is assumed either of the order of the experimental energy resolution ($\Delta=1.0$
MeV, first and second columns), or smaller ($\Delta=0.2$ MeV, third column). The experimental data~\cite{Wakasa2012} are displayed by black dots and black solid lines. } \label{fig1}
\end{figure*}
%----------------------------------------------------------------------------------------------------------

In Fig. \ref{fig1}, we show the GT strength distributions,
their cumulative sums and the scaled cumulative sums calculated with the Skyrme interactions SAMi and SGII
for the nucleus $^{208}$Pb, compared with the experimental data of Ref. \cite{Wakasa2012}. The excitation
energies are given with respect to the parent nucleus in all the figures. The discrete RPA strength is folded with
Lorentzian functions having width equal to $2\Delta$. In order to obtain a consistent comparison with data,
in panels (a)-(f) we adopt a value $\Delta=1$ MeV, similar to the experimental energy resolution, in both
the RPA and RPA+PVC calculations. In panels (g)-(i) we also show the results obtained with a smaller value,
namely $\Delta=200$ keV, in order to
see the features of the theoretical GT distribution in more detail. The scaled cumulative sums are obtained
by scaling both the RPA and RPA+PVC total strength at $E=25$ MeV to the experimental value.

As already seen in Fig. \ref{fig01}, the Skyrme interaction SAMi reproduces the experimental
energy of the GTR very well at the RPA level.
Including the coupling with phonons, the GT energy is shifted downward, worsening the agreement with the experimental peak energy. On the other hand, the SGII interaction produces a higher GT peak energy compared to the experiment at the RPA level, while the RPA+PVC calculation reproduces very well the line shape of the resonance. Our results are
similar to those obtained in the recent relativistic time blocking (RTBA) calculations
for $^{208}$Pb reported in Ref.~\cite{Litvinova2014}. Besides the main GT peak at 19.2 MeV, there is another low-energy peak produced by the RPA+PVC calculations, located at about 11.5 MeV (SAMi) or 12.5 MeV (SGII).

Concerning strengths, the total GT$^-$ [$\sum B$(GT$^-$)] and GT$^+$ [$\sum B$(GT$^+$)] strengths calculated in the RPA approach with the SGII interaction are 132.95 and 0.96, respectively. The RPA result for $\sum B({\rm GT}^-)-\sum B({\rm GT}^+)$ exhausts 99.99\% of the Ikeda sum rule. Only 3\% of the calculated GT$^-$ strength lies at energies above 25 MeV. In the RPA+PVC calculation,
$\sum B({\rm GT}^-)-\sum B({\rm GT}^+)$ exhausts 95.2\% of the Ikeda sum rule up to the excitation energy 60 MeV while this value becomes 97.3\% in the case of smearing parameter $\Delta=0.2$ MeV. About 15\% of the sum rule is shifted above $E=25$ MeV
[$\sum_{E \leq  25 {\rm MeV}} B({\rm GT}^-) = 112.48$, $\sum_{E \leq 25 {\rm MeV}} B({\rm GT}^+) = 0.72$]. The experimental strength integrated up to 25 MeV is equal to about 79, corresponding to $71\%$ of the RPA+PVC result
\cite{Wakasa2012}. This is in agreement with the recent RTBA calculation
of Ref. \cite{Litvinova2014}, where the ratio between the experimental
strength integrated up to 25 MeV and the RTBA result is about 72\%
using the same smearing parameter $\Delta=1.0$ MeV as ours.
Previous studies found that RPA tensor correlations could
shift about 10\% of the sum rule to the excitation energy
region above 30 MeV  \cite{Bai2009,Bai2009a}.
The inclusion of $\Delta$-isobar excitation  could
move strength to very high excitation energy, this amount
being of the order
of 10\% of the total sum rule or less \cite{Brown:1987,Wakasa:1997,Yako:2005}.
Concerning the remaining discrepancy with experiment, we cannot determine
to which extent it can be attributed to deficiencies in the model, or to systematic uncertainties that the experiment was unable to pin down.
Despite the disagreement in the total strength, the energy dependence of the cumulative sum
calculated with RPA+PVC reproduces quite well the experimental data.
In order to see this more clearly, in panel (f) we scale the theoretical integrated strength up to $E=25$ MeV
to the experimental value. It can be seen that the inclusion of the phonon coupling improves the description of data
considerably as compared to the RPA result. There is some excess in the theoretical strength of the
RPA+PVC calculation as compared to the data in the energy region $E=12-16$ MeV, due to the already mentioned
low-lying peaks.

By reducing the value of the smearing parameter from $\Delta =1$ MeV to $\Delta = 200$ keV [panel (g)],
one can investigate the detailed structure of the resonance.
The main peak, which had a FWHM equal to 3.6 MeV (cf. Fig. \ref{fig01}
and Table \ref{table0}), is roughly split into two peaks, located at $E=$ 19.2 MeV and $E= 20.6$ MeV, with a FWHM equal to 1.2 and 2 MeV respectively (cf. Table \ref{table0}). In Ref. \cite{Litvinova2014}, the resonance calculated in RTBA is also split into two subpeaks in a very similar way.

%---------------------------------------------------------------------------------------------------------
\begin{figure*}
\includegraphics[scale=0.2,angle=0]{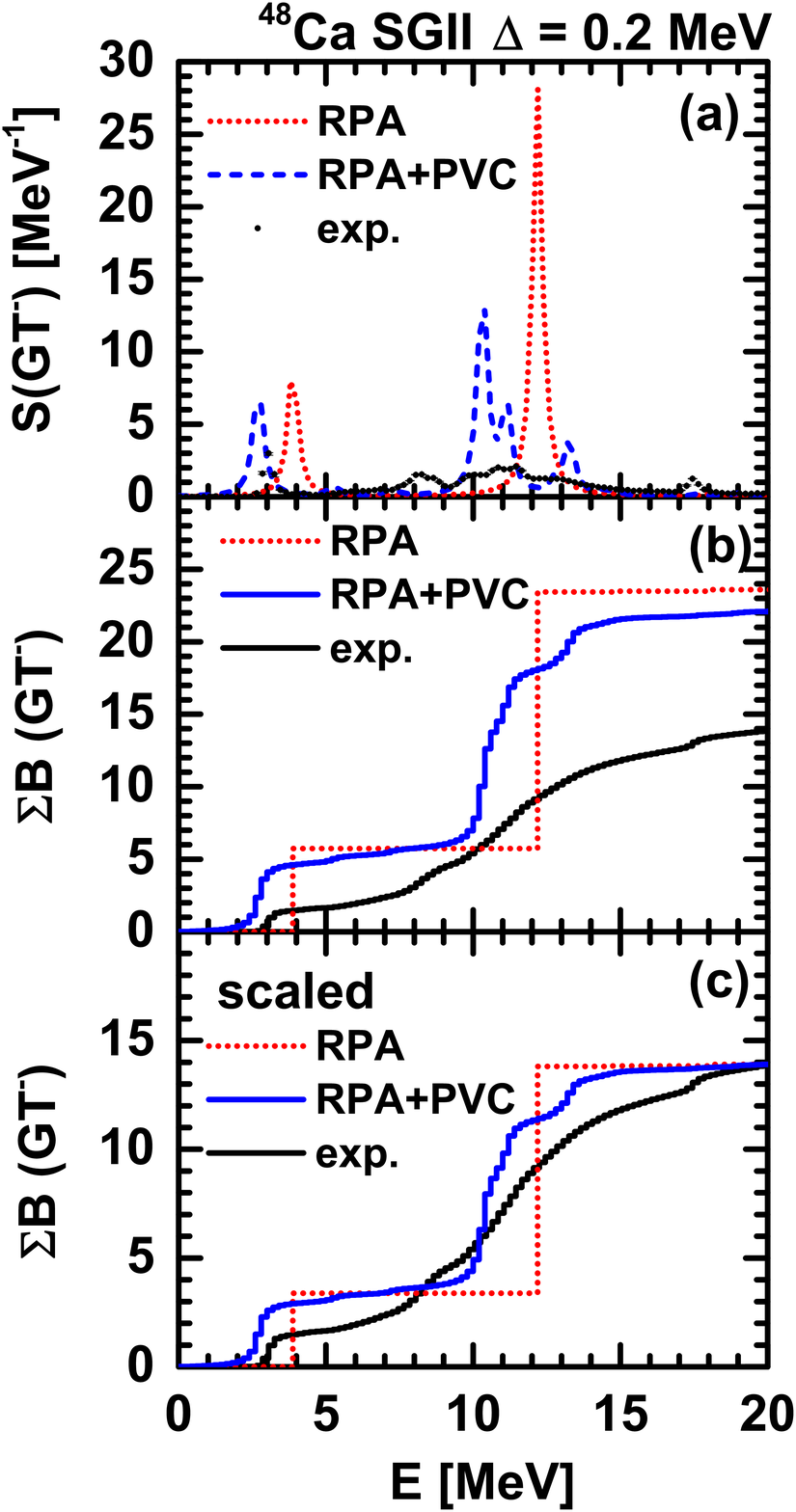}
\includegraphics[scale=0.2,angle=0]{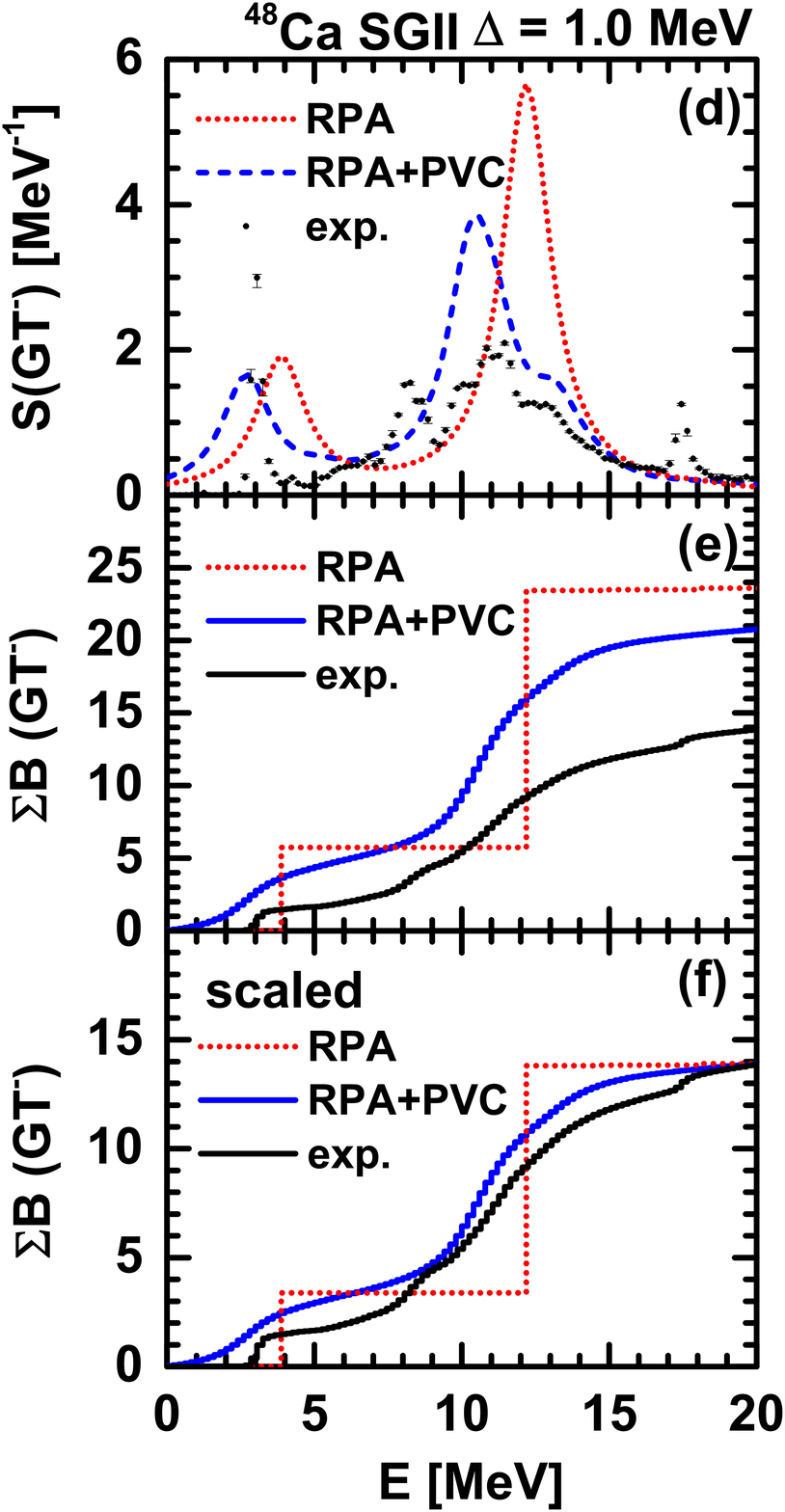}
 \caption{(Color online) The same as Fig. \ref{fig1}, in the case of the nucleus $^{48}$Ca and the interaction SGII.
The experimental data are taken from Ref. \cite{Yako2009}.} \label{fig3}
\end{figure*}
%----------------------------------------------------------------------------------------------------------
In keeping with the fact that the effects of PVC depend little  on the interaction, in the following we will only display the results calculated with the interaction SGII.
In Fig. \ref{fig3}, we show GT results  for  $^{48}$Ca.  In this case,
the experimental energy resolution is about 200 keV,  and
the experimental strength function  \cite{Yako2009} displays a rather  complex structure.
One finds a low-lying  peak at about 3 MeV, with a narrow FHWM of  about 0.4 MeV,
followed by a broad resonance region between 5 and 16 MeV displaying
two peaks lying at 8.2 MeV (with a FWHM of 1.5 MeV)  and at 10.8 MeV (with a FWHM of 3.9 MeV);
the  centroid energy of these two peaks is equal to  10.5 MeV.
Finally, a  small and narrow peak is observed at 17.5 MeV. The RPA+PVC calculation
with the small averaging parameter $\Delta = 0.2 $ MeV reproduces the strength distribution
in the low-energy region reasonably well: the lowest  peak energy (2.8 MeV),  as well as its FWHM  (0.4 MeV),
match the experimental values.
One finds, then, a very large peak with centroid energy at 10.5 MeV and a much smaller peak at 13.2 MeV.
The associated FWHMs are too narrow, being equal to 1.2 MeV and 0.7 MeV  respectively (cf. Table \ref{table0}).
The calculation with $\Delta$ = 1 MeV, in which these two peaks merge in a single peak with
centroid energy 10.4 MeV and FWHM 2.6 MeV,
provides a better overall   description of the experimental line shape [cf. panel (d)],  as well as a better
reproduction  of the observed cumulated strength  once it is suitably scaled [cf. panel (f)].
The peak appearing at 17.5 MeV in the experiment is not reproduced by the calculation.
With $\Delta=0.2$ MeV, the experimental strength integrated up to 20 MeV  reaches 63\% of the RPA+PVC result. In turn, 8\% of the
sum rule in the  RPA+PVC calculation (2\% in the RPA case)  is found beyond this interval.

%---------------------------------------------------------------------------------------------------------
\begin{figure*}
\includegraphics[scale=0.2,angle=0]{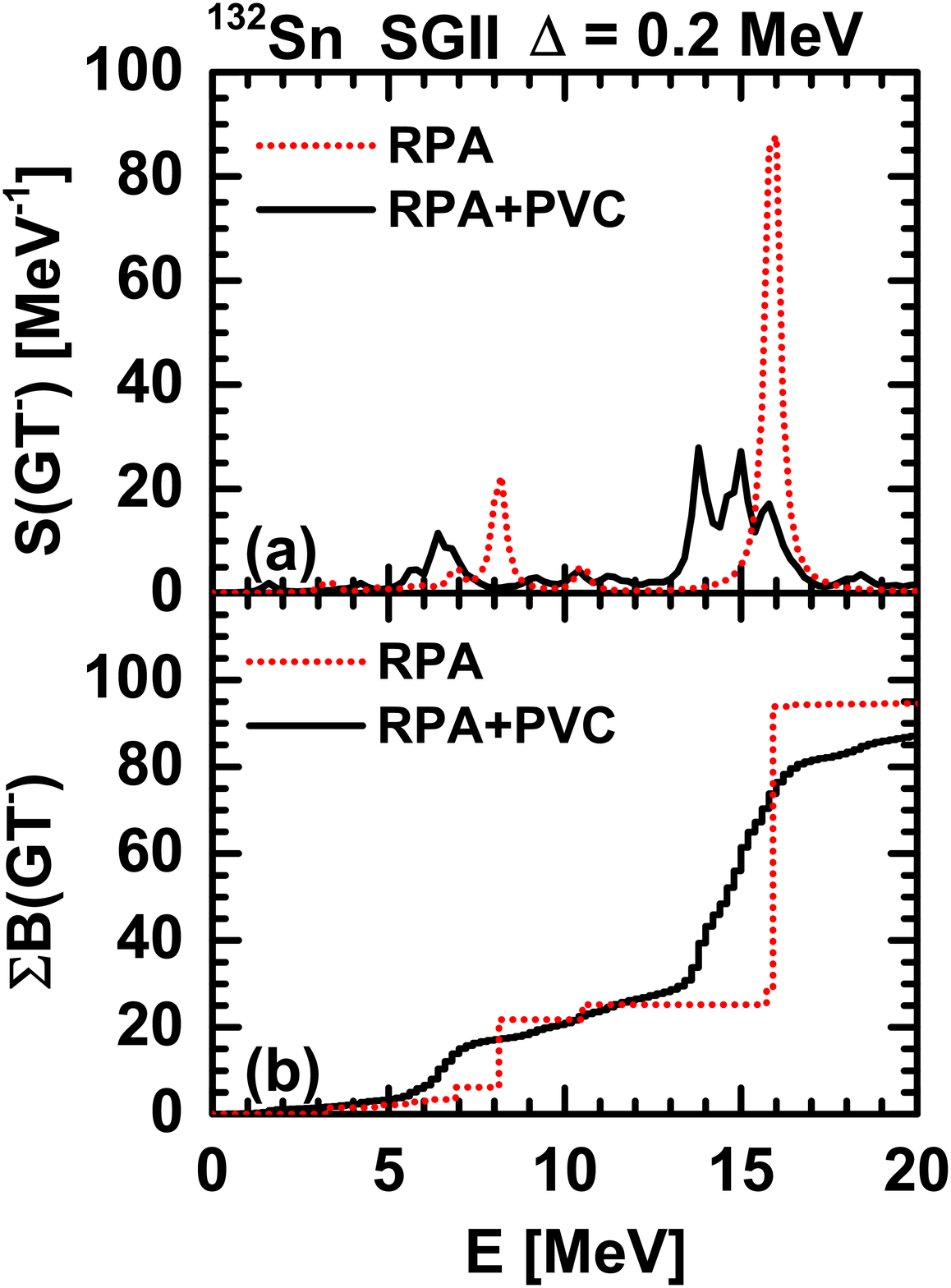}
\includegraphics[scale=0.2,angle=0]{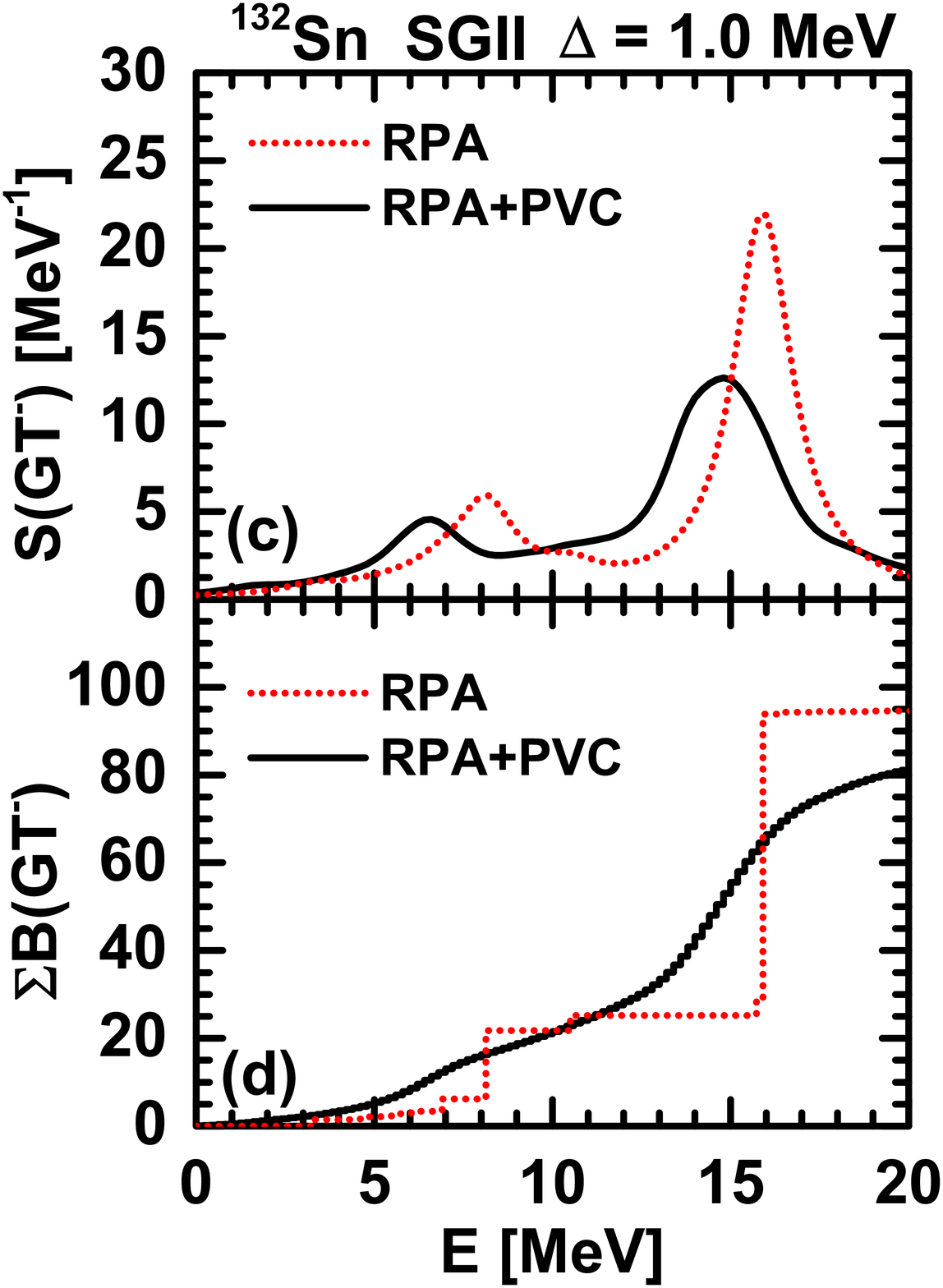}
 \caption{(Color online) Gamow-Teller strength distributions [panels (a) and (c)] with their cumulative sums
[panels (b) and (d)] calculated with the Skyrme interaction SGII for the nucleus $^{132}$Sn.
The red dotted lines and black solid lines denote the results of the RPA and RPA+PVC models, respectively.
The smearing parameter $\Delta$ in the calculations takes either a small value $\Delta=0.2$ MeV (first column)
or a large one $\Delta=1.0$ MeV (second column). } \label{fig4}
\end{figure*}
%----------------------------------------------------------------------------------------------------------

In the following, we will provide predictions for a few nuclei for which no measurement has been published up to date.

In Fig. \ref{fig4} we show the GT strength distributions and their cumulative sums for the nucleus $^{132}$Sn. Very recently, a  (p,n) experiment
on $^{132}$Sn was carried out at RIKEN with the purpose of studying its GT and spin dipole strength distributions \cite{Sasano2014}.
In our calculation, using  the small averaging  parameter $\Delta=0.2$ MeV, the  PVC lowers the main RPA peak
located at 16 MeV  by about 2 MeV and fragments it into three close
sub-peaks, which, by  using $\Delta= 1$ MeV, merge into one broad peak with a FWHM of about 3.6 MeV in the resonance
region (cf. Table \ref{table0}). A secondary RPA peak at about 8 MeV is also shifted downward by 2 MeV.
About 1.5\%  (RPA) and 9\% (RPA+PVC with $\Delta$ =0.2 MeV) of the cumulative sum rule is  found beyond 20 MeV.
The latter value increases up to 15\% using $\Delta$ =1 MeV.

%---------------------------------------------------------------------------------------------------------
\begin{figure*}
\includegraphics[scale=0.2,angle=0]{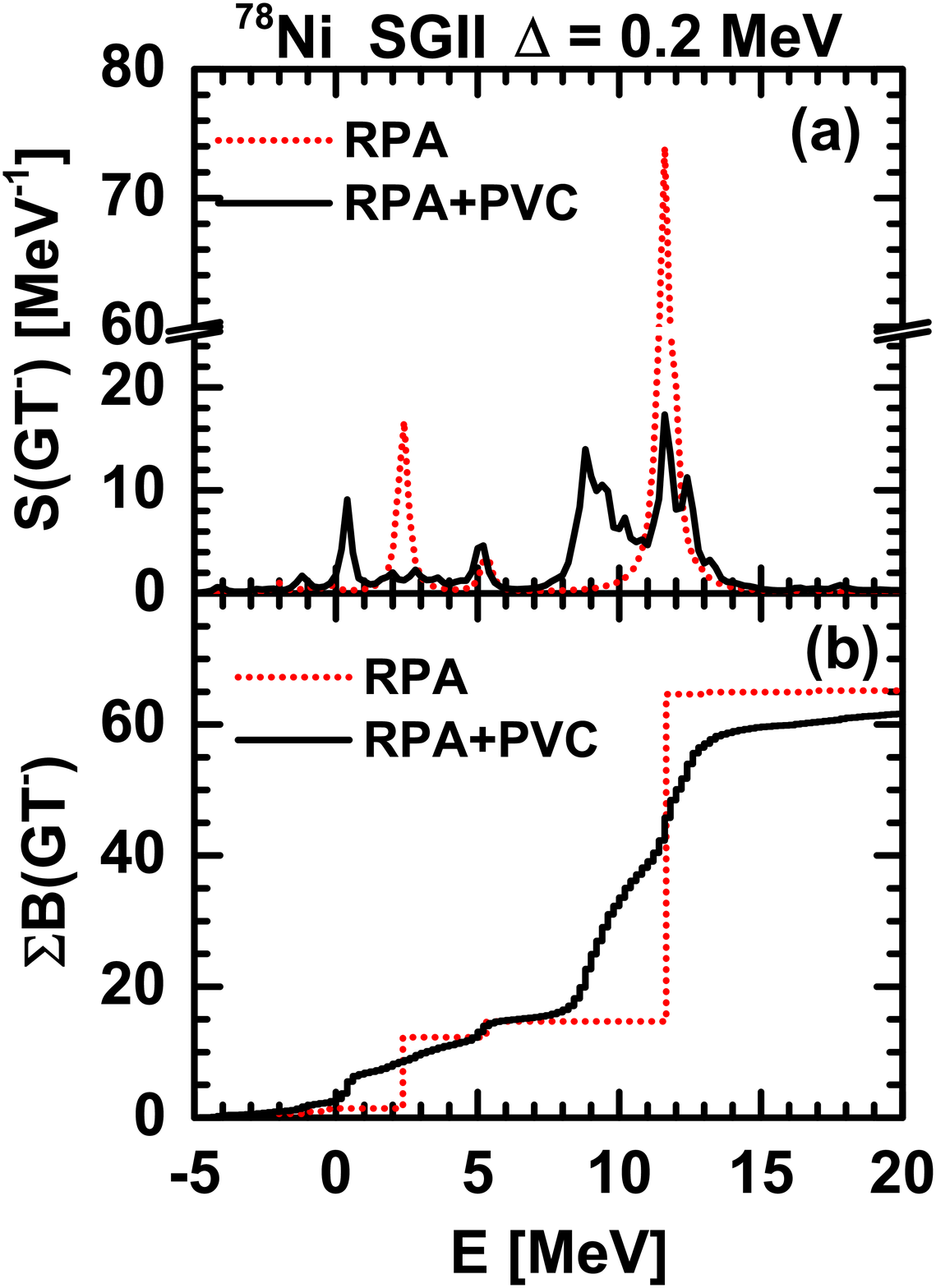}
\includegraphics[scale=0.2,angle=0]{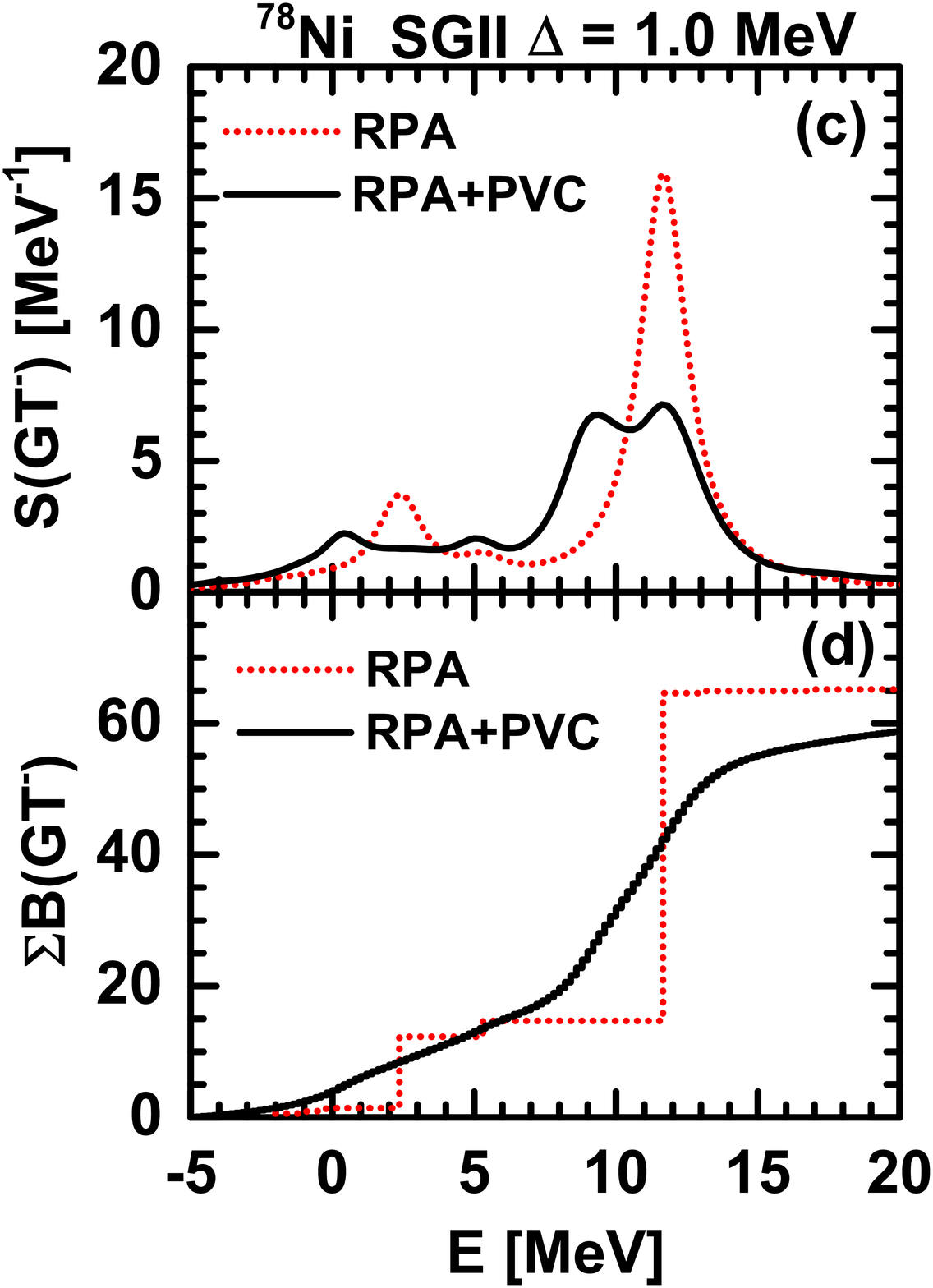}
 \caption{(Color online) The same as Fig. \ref{fig4} in the case of the nucleus $^{78}$Ni.} \label{fig6}
\end{figure*}
%----------------------------------------------------------------------------------------------------------

Finally, in Fig. \ref{fig6} the strength distribution calculated  for the nucleus $^{78}$Ni is shown.
This  nucleus lies on the r-process path and its $\beta$-decay half-life has a considerable  influence
on the nuclear abundances around $N=50$ \cite{Hosmer2005}.
Its $\beta$-decay half-life has been measured \cite{Hosmer2005},
and it is quite short ($\approx$ 110 ms) as expected from the large
neutron excess so that the measurement of the GT strength distribution definitely represents  a very severe challenge.
In our calculation, the inclusion of the particle-vibration coupling (calculated with $\Delta = 0.2$ MeV)
produces  in this case a large spreading width and a strong fragmentation of the
main  RPA resonance peak located at about 11.5 MeV,
leading to a two-peak structure (cf. Table \ref{table0}) centered at 10.7 MeV with a FWHM of about 4.2 MeV.
Also the low-lying RPA peak is shifted downward by about 2 MeV and becomes broader.
These effects on the low-lying peak could
help to reduce the calculated $\beta$-decay half-life which is usually overestimated in the RPA approach \cite{Engel1999,Niksic2005,Niu2013}.
As for the cumulative sum, 1.3\% of the RPA  sum rule and 7\% of the RPA+PVC sum rule using $\Delta$ =0.2 MeV (11\% using $\Delta$ =1 MeV) are found  beyond $E=20$ MeV.

\subsection{Overall features of Gamow-Teller resonance and phonons}
\label{discussion1.5}

%---------------------------------------------------------------------------------------
\begin{table}
\caption{Summary of the GTR peak energies and their associated FWHMs extracted from  the
strength functions of the nuclei we have considered, calculated by the RPA+PVC approach
with the interaction SGII. Also shown are available experimental data.}
\begin{tabular}{cccccc}
\hline\hline
            &           &  $^{208}$Pb \quad  \quad &  $^{48}$Ca \quad \quad  & $^{132}$Sn\quad  \quad &    $^{78}$Ni \quad\quad \\
\hline
  Expt.      & $E_1$     &  19.2& 8.3 & &  \\
            & FWHM$_1$  &   5.0&  1.5 & &  \\
            & $E_2$     &      &  10.9 & &  \\
            & FWHM$_2$  &      &  3.9 & & \\
            \hline
$\Delta=1$ MeV & $E_1$  & 19.6 &  10.4 &  14.8 &   10.5\\
            & FWHM$_1$  & 3.6  &  2.6  &  3.6 &  5.6 \\
 \hline
$\Delta=0.2$ MeV & $E_1$&  19.2&  10.4  &  13.8 &  8.8 \\
            & FWHM$_1$  &  1.2 &  0.5   &  0.7  &  1.8  \\
            & $E_2$     &  20.6&  11.2  & 15    &  11.6\\
            & FWHM$_2$  &  2.0 &  0.7   & 0.9   &  1.4 \\
            & $E_3$     &      &  13.2  & 15.8  &   \\
            & FWHM$_3$  &      &  0.7   & 1.0   &     \\
\hline \hline
\end{tabular}
\label{table0}
\end{table}
%--------------------------------------------------------------------------------------------

In the previous subsection, we have seen that the coupling  to vibrations
modifies the strength distribution in rather different ways, depending on the
specific nucleus.
To summarize this feature,
the energies of the GTR peaks and the associated FWHMs taken from  the
strength functions calculated by the  RPA+PVC approach with the interaction SGII,
are provided in Table \ref{table0}. Also the available experimental data are shown.
Using the value $\Delta= 0.2 $ MeV, one finds two or three peaks  with an overall FWHM of
2 MeV in $^{48}$Ca, and 3 MeV in  $^{208}$Pb and in $^{132}$Sn,
and  a broad and strongly fragmented structure in $^{78}$Ni.

%--------------------------------------------------------------------------------------------
\begin{table}
\caption{The imaginary part of the self-energy multiplied by $-2$ calculated at the RPA  energy of the GTR peak
in the case of the
nuclei $^{208}$Pb, $^{48}$Ca, $^{132}$Sn and $^{78}$Ni with the interaction SGII, using the
smearing parameter $\Delta=0.2$ MeV. The partial contributions
from phonons with different multipolarities, as well as the total values,
are listed.}
\begin{tabular}{ccccccc}
  \hline\hline
   Nucleus &  $-2 {\rm Im}\Sigma (1^-)$ &  $-2 {\rm Im}\Sigma  (2^+)$ &  $-2 {\rm Im}\Sigma  (3^-)$ &  $-2 {\rm Im}\Sigma  (4^+)$ &  $-2 {\rm Im}\Sigma  (5^-)$ & $-2 {\rm Im}\Sigma $ (total) \\
   \hline
   $^{208}$Pb & 0.54 & 0.075 &  0.89 &0.078 & 0.29 & 1.88 \\
   $^{48}$Ca  & 1.0 $\times 10^{-3}$ & 1.75 & 0.016 & 0.20 & 6.6 $\times 10^{-4}$& 1.96\\
   $^{132}$Sn & 0.67& 0.30& 0.046 & 0.40 & 0.030 & 1.46 \\
   $^{78}$Ni  & 0.37 & 0.17 & 0.22 &0.56 & $3.1 \times  10^{-3}$  & 1.32 \\
   \hline\hline
\end{tabular}
\label{table3}
\end{table}
%-----------------------------------------------------------------------------------------------

In Table \ref{table3},  we provide approximate values
for the widths of the various nuclei, calculated according to
the perturbative expression  $\Gamma_{\rm GTR}(E_{\rm RPA})$
[cf. Eq. (\ref{Gamma})] with $\Delta$= 0.2 MeV. The partial contributions
from phonons with different multipolarities are also listed. These values
are approximated ones, as discussed below in the cases of $^{48}$Ca and $^{208}$Pb.
However, they provide useful information about the differences and similarities in widths among these nuclei.
The $2^+$ phonon gives the dominant
contribution  to the width in $^{48}$Ca,  while only negative-parity phonons are important for $^{208}$Pb. The $5^-$
multipolarity  is relevant only in $^{208}$Pb. The other phonons give comparable contributions in $^{78}$Ni and $^{132}$Sn (except  for $3^-$ in $^{132}$Sn).

%-----------------------------------------------------------------------------------------------
\begin{table}
\caption{The energies and reduced transition probabilities of the lowest phonons of
multipolarity $2^+, 3^-, 4^+$, and $5^-$ calculated in the RPA with interaction SGII are compared with available experimental data,
for the nuclei $^{208}$Pb, $^{48}$Ca, $^{132}$Sn, and $^{78}$Ni.
The experimental data are taken from Refs. \cite{nndc,Martin2007,Khazov2005,Burrows2006}.
The calculated phonon states shown in parentheses have a strength smaller than 5\% of the total isoscalar or isovector strength, and therefore are not included in the PVC calculation.
}
    \scriptsize{
  \begin{tabular}{cclllll}
    \hline\hline
            & & &  $^{208}$Pb &  $^{48}$Ca & $^{132}$Sn &  $^{78}$Ni  \\
            \hline
    $2^+$   & Expt. &  $E$ (MeV) & 4.09  & 3.83 & 4.04 &  \\
            &      &  $B(EL,0\rightarrow L)$(e$^2$ fm$^{4}$) &  $3.10 \times 10^3$ &  88.84 & $1.39 \times 10^3$ &  \\
            &      & $B(EL)$ (s.p.u.) & 8.5 & 1.7 & 7.0 & \\
            & Theor.&  $E$ (MeV) & 5.03  & 3.80 & 4.54 &  3.46 \\
            &      &  $B(EL,0\rightarrow L)$(e$^2$ fm$^{4}$) & $2.74\times 10^3$ &  51.82 & $1.10 \times 10^3$ & $3.83 \times 10^2$ \\
            &      & $B(EL)$ (s.p.u.) & 7.5 & 1.0 & 5.5 & 3.9\\
    \hline
    $3^-$   & Expt. &  $E$ (MeV) & 2.62  & 4.51 & 4.35 &     \\
            &      &  $B(EL,0\rightarrow L)$(e$^2$ fm$^{6}$) & $6.12\times 10^5$ &$4.82\times 10^3$& $>5.14\times 10^4$   &    \\
            &      & $B(EL)$ (s.p.u.) & 34.1  & 5.0 & $>7.1$ & \\
            & Theor.&  $E$ (MeV) & 3.09  & 5.75 & 5.29 &  7.61 \\
            &      &  $B(EL,0\rightarrow L)$(e$^2$ fm$^{6}$) & $6.79\times 10^5$  & $8.02\times 10^3$ & $1.19 \times 10^5$ & $1.68 \times 10^4$ \\
            &      & $B(EL)$ (s.p.u.) & 37.7  & 8.4 & 16.5 & 6.6 \\
    \hline
    $4^+$   & Expt. &  $E$ (MeV) & 4.32  & 4.50 & 4.42 &  \\
            &      &  $B(EL,0\rightarrow L)$(e$^2$ fm$^{8}$) & $1.62\times 10^7$  &   &$2.17\times 10^6$ &  \\
            &      & $B(EL)$ (s.p.u.) & 18.8 & & 8.0 & \\
            & Theor.&  $E$ (MeV) & 4.96  & (4.13)  & 5.00 &  4.22 \\
            &      &  $B(EL,0\rightarrow L)$(e$^2$ fm$^{8}$) & $9.77\times 10^6$  & ($6.82\times 10^3$)  & $2.08 \times 10^6$ &  $2.12 \times 10^5$ \\
          &      & $B(EL)$ (s.p.u.) & 11.4  & (0.4) & 8.1 & 3.4 \\
    \hline
    $5^-$   & Expt. &  $E$ (MeV) & 3.20  & 5.73 & 4.94 &  \\
            &      &  $B(EL,0\rightarrow L)$(e$^2$ fm$^{10}$) & $4.47 \times 10^8$    &  &  &   \\
            &      & $B(EL)$ (s.p.u.) &11.0 & & & \\
            & Theor.&  $E$ (MeV) & 3.76  & (7.28)   & 6.85 &  (8.11) \\
            &      &  $B(EL,0\rightarrow L)$(e$^2$ fm$^{10}$) & $4.64\times 10^8$   & ($1.58\times 10^6$)   & $2.43 \times 10^7$ & ($2.61 \times 10^3$) \\
         &      & $B(EL)$ (s.p.u.) & 11.4  & (5.1) & 2.7 & (0.0017) \\
            \hline\hline
  \end{tabular}
  }

  \label{tablephonon}
\end{table}
%---------------------------------------------------------------------------------------------------

These different behaviors are  mostly determined by the properties of the
lowest phonons  and  by the
underlying shell structure in each nucleus.
In Table \ref{tablephonon} we report
the energies and reduced transition probabilities of the lowest phonons with
multipolarity $2^+, 3^-, 4^+$, and $5^-$  calculated within RPA for all these nuclei,
comparing them  with  available experimental data.
In most cases, the states have rather collective character, their reduced transition probabilities being of the order of several single-particle units (s.p.u.).
For $^{208}$Pb and $^{132}$Sn, both the energies and reduced
transition probabilities agree well with the experimental data.
In the case of $^{48}$Ca, although the calculated phonon energies
agree well with the data, the theoretical and experimental reduced
transition probabilities differ by about 40\%: theory underestimates
the B(E2) value and overestimates the B(E3) value.
Since among the 2$^+$ phonons that provide a dominant
contribution to the width, the lowest 2$^+$ state
plays a major role, we can argue that the calculated
width would be approximately doubled if the B(E2) value were enlarged
by the same factor,
in which case the experimental width would be well reproduced.
Thus, the calculated width
might be underestimated not due to  a basic failure in our picture
but rather to the fact that RPA does not account well in this
case for the collectivity of the low-lying 2$^+$ phonon.

\subsection{Underlying mechanisms for the spreading width and fragmentation}
\label{discussion2}

%---------------------------------------------------------------------------------------------------------
\begin{figure*}
\includegraphics[scale=0.3,angle=0]{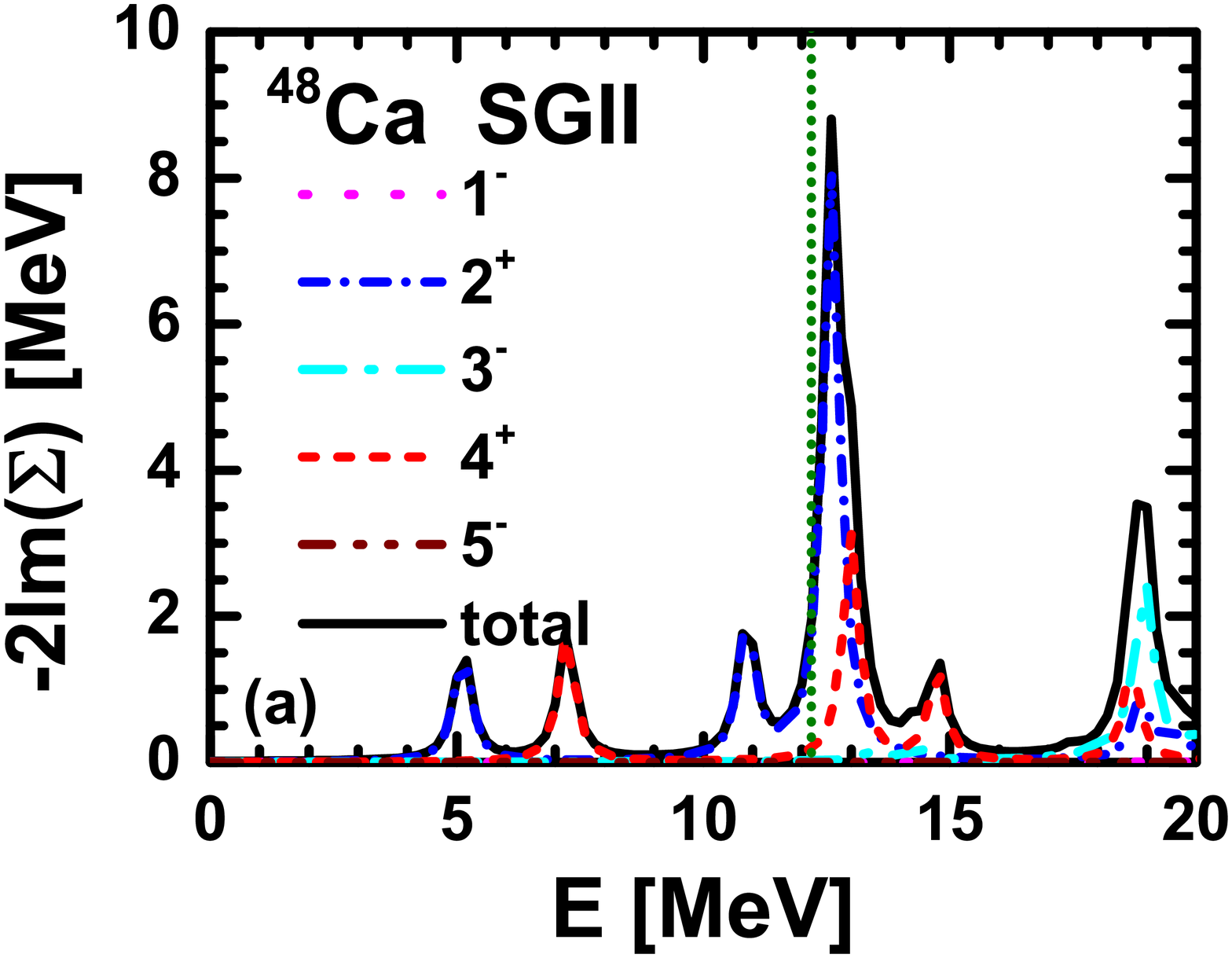}
\includegraphics[scale=0.3,angle=0]{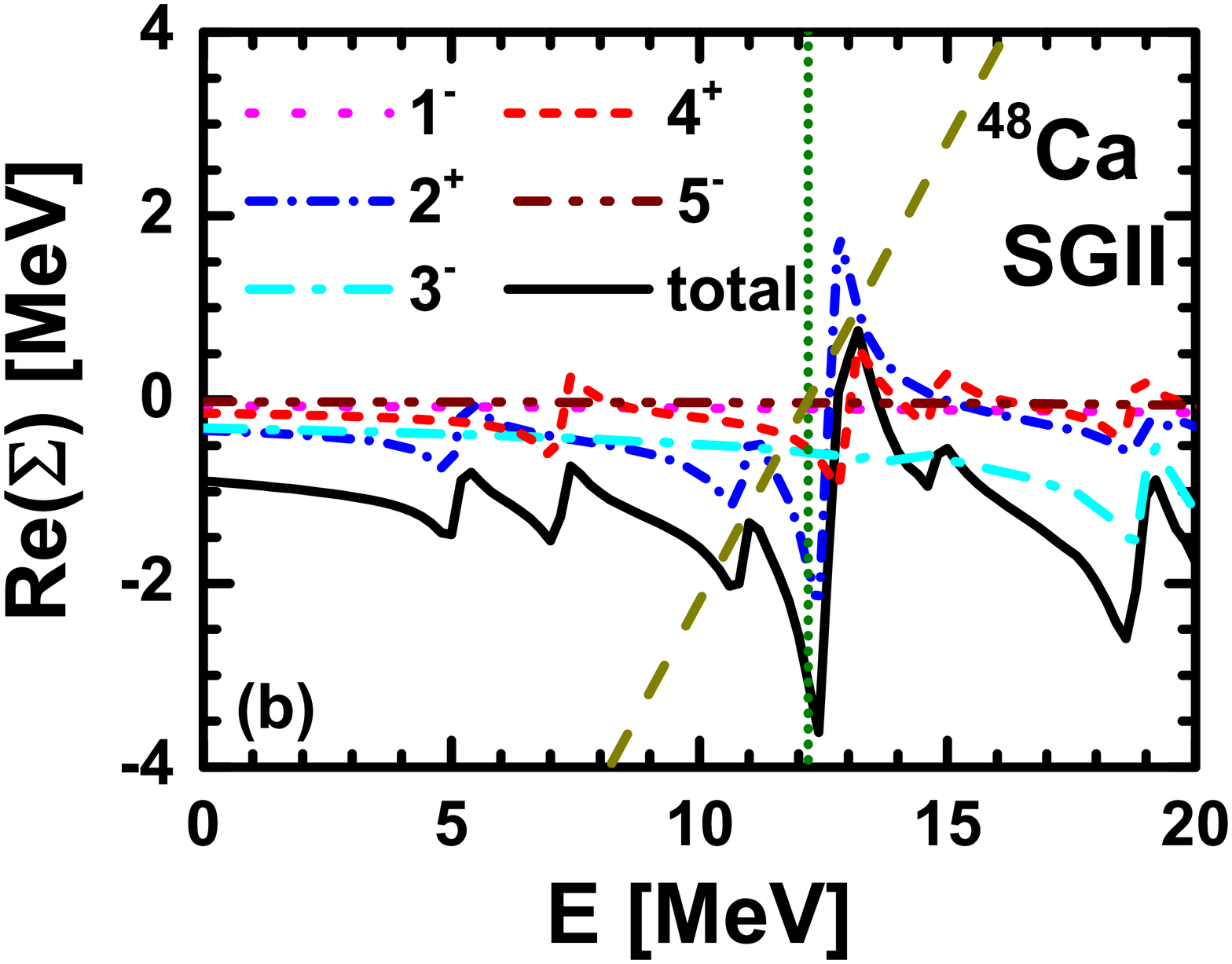}
\caption{(Color online) The imaginary part multiplied by $-2$ [panel (a)] and
the real part [panel(b)] of the self-energy of the main RPA state for the nucleus
$^{48}$Ca, calculated by the interaction SGII with a smearing parameter $\Delta=0.2$ MeV.
In addition to the total results, the contributions from phonons with different
multipolarities are shown separately. The vertical short-dotted-olive line
represents the energy position $E_{\rm RPA}$ of the GTR peak calculated in the RPA approach,
while the dashed line in panel (b) represents the function $y=E-E_{\rm RPA}$.} \label{fig7}
\end{figure*}
%---------------------------------------------------------------------------------------------------------

%---------------------------------------------------------------------------------------------------------
\begin{figure*}
\includegraphics[scale=0.3,angle=0]{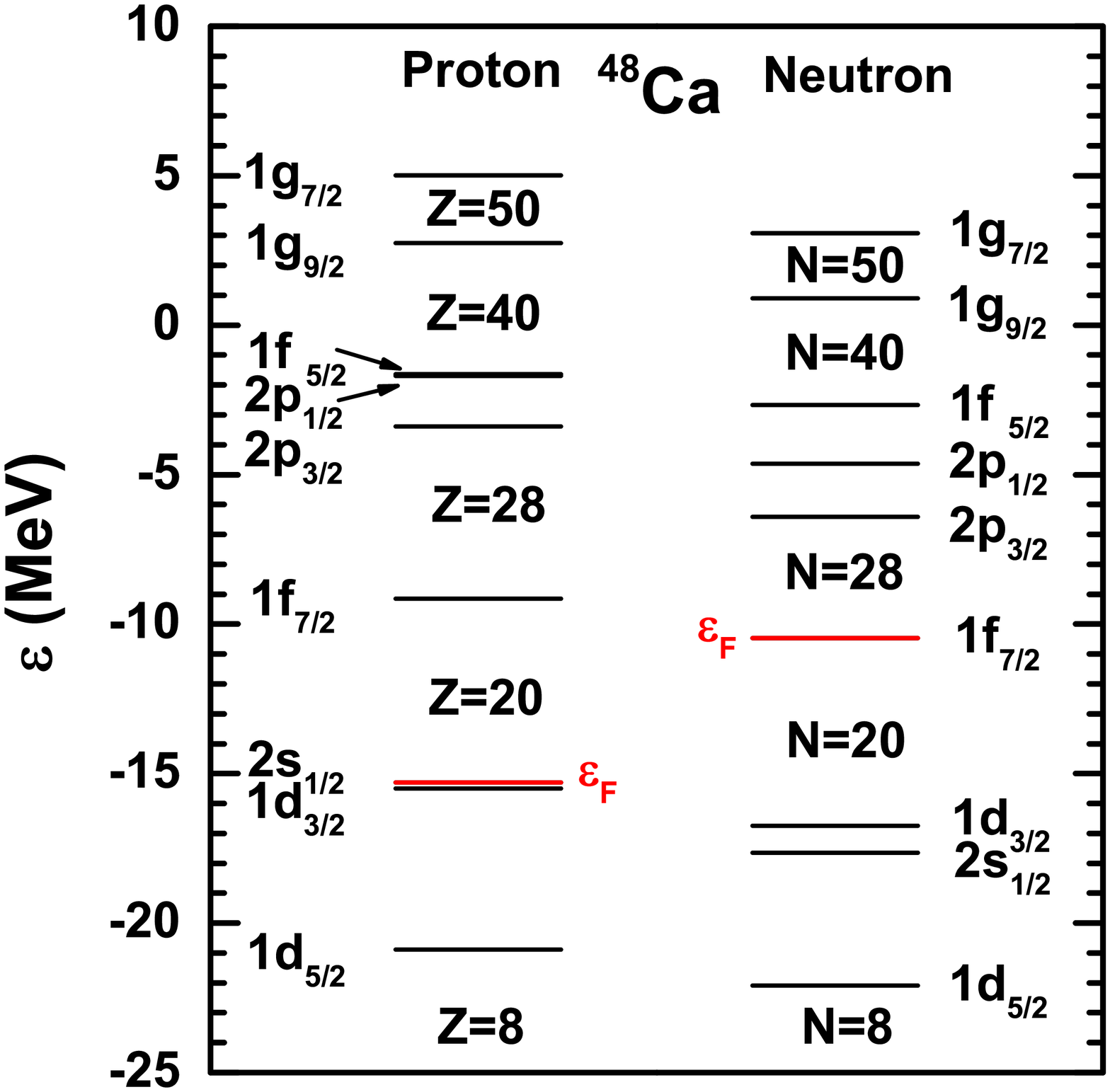}
 \caption{(Color online)  Proton and neutron single-particle spectrum in $^{48}$Ca obtained from the HF calculation with SGII interaction.}
 \label{Ca48SP}
\end{figure*}
%---------------------------------------------------------------------------------------------------------

In this subsection
we will discuss in more detail the microscopic processes
leading to the damping of the GT resonance.
We start by considering the case of $^{48}$Ca, calculated with $\Delta$ = 0.2 MeV.
Looking at Tables \ref{table0} and \ref{table3} and Fig. \ref{fig7}, we observe  that

(i) The calculated strength function displays three peaks located at 10.4, 11.2 and 13.2 MeV,
with associated widths equal to  0.5, 0.7,  and 0.7 MeV.

(ii) The dominant contribution to the width
is provided by the coupling to quadrupole phonons;
 $4^+$ phonons become important beyond 13 MeV (cf. Fig. \ref{fig7}).
 The contributions from negative parity phonons are negligible.

The result (i)  obtained in the full diagonalization can be quantitatively reproduced by the
much simpler expressions with diagonal approximation
Eqs. (\ref{Omega}) and (\ref{Gamma}),  in which
the self-energy $\Sigma_{\rm GTR}(\omega)$  for the giant resonance state
is given by  the energy-dependent matrix element  $({\cal A}_1)_{\rm GTR,GTR}(E)$.
For convenience we reproduce the  equations here, in a slightly different form:
\begin{equation}
E_{\rm GTR} -  E_{\rm RPA} = {\rm Re}[({\cal A}_1)_{\rm GTR,GTR}(E_{\rm GTR})]
\label{Omega1}
\end{equation}
and
\begin{equation}
 \Gamma_{\rm GTR}(E_{\rm GTR})= -2{\rm Im}[({\cal A}_1)_{\rm GTR, GTR}(E_{\rm GTR})].
\label{Gamma1}
\end{equation}

The  curve  $ {\rm Re}[({\cal A}_1)_{\rm GTR,GTR}(E)]$ and the
line $y=E - E_{\rm RPA}$ are plotted in panel (b) of Fig. \ref{fig7}. They cross at
$E$= 10.4 MeV. Furthermore, the quantity $\vert y - {\rm Re}[\Sigma_{\rm GTR}(E)]\vert$
has two minima very close to 0 at $E= 11.2$ and $13.2$ MeV. These values give the GTR peak energies $E_{\rm GTR}$. Correspondingly, the widths $ \Gamma_{\rm GTR}(E_{\rm GTR})$ for $E_{GTR}$ = 10.4,11.2, and 13.2 are given by  0.40, 0.79, and 2.52 MeV  , respectively.
 These values  of the energies and widths
are in very good agreement with  the complex  eigenvalues resulting from the
full solution which are   $10.30 - i 0.41$ MeV,
$10.76 -i 0.83$ MeV, and $12.96-i 2.36$ MeV, where the imaginary parts have been multiplied by 2 to obtain the width.
The first two values correspond well to the FWHMs (0.5 and 0.7 MeV) extracted from the strength distribution (cf. Table \ref{table0}).
However, the FWHM of the third peak at $13.2$ MeV obtained from the strength function is 0.7 MeV, smaller than the value given above due to the sharp decrease of
the imaginary part of the self-energy above the peak energy [cf. panel (a) of Fig. \ref{fig7}].

In order to understand the feature (ii), we need to determine the configurations which give the largest contributions
to the particle and hole self-energies, i.e., diagrams (1) and (2) of Fig. \ref{fig0}.
The microscopic RPA wave function of the GTR in $^{48}$Ca is dominated by  a single p-h transition of energy 8.84 MeV,
namely $\nu 1{\rm f}_{7/2} \rightarrow \pi 1{\rm f}_{5/2}$. In diagram (1) (cf. Fig. \ref{fig0}), the most important  intermediate proton particle states $p''$
and phonons $nL$
are those being able to couple with the $\pi 1{\rm f}_{5/2}$ proton state and  minimize the denominator
$\Omega_{\rm GTR} -(\omega_{nL}+\epsilon_{\pi p''}-\epsilon_{\nu 1{\rm f}_{7/2}}) +i\Delta $
in Eq. (\ref{Wdown1}).  The GTR energy is given approximately by the
energy of the particle-hole transition, plus a shift $\Delta E $, which takes into account the effects
of  the  repulsive p-h interaction and  of the PVC (cf. Fig. \ref{fig01}).
When we use a small value of the smearing parameter $\Delta$, several peaks may appear in the strength distribution (cf. Fig. \ref{fig3}). For simplicity, in the following approximate analysis, we shall use the RPA energy $E_{\rm RPA}$ instead of $\Omega_{\rm GTR}$, and neglect the PVC effect on $\Delta E$.
We then put  $E_{\rm RPA} =  \epsilon_{\pi 1{\rm f}_{5/2}} -
\epsilon_{\nu 1{\rm f}_{7/2}}+ \Delta E$, so that the denominator becomes
$\epsilon_{\pi 1{\rm f}_{5/2}} - \epsilon_{\pi p''}- \omega_{nL} + \Delta E +i\Delta $.
This means that the relevant intermediate states must lie at an  energy
$\epsilon_{\pi p''}  \approx   \epsilon_{\pi 1{\rm f}_{5/2}} - \omega_{nL} + \Delta  E $.
Given  that typical values for the energies of the low-lying collective phonons
$\omega_{nL}$ are 3-4  MeV and $\Delta E =3$ MeV, this requires that $p''$ lies close to $\pi 1{\rm f}_{5/2}$.
Looking at Fig. \ref{Ca48SP}, one realizes that this condition is
fulfilled only by  negative parity single-particle levels
in the $pf$ shell, which can couple to the state $ \pi 1{\rm f}_{5/2}$ only through  positive parity phonons.
In a similar way,  for diagram (2) (cf. Fig. \ref{fig0}) the energy of the intermediate neutron  hole states $\nu h''$ coupling to the
$\nu1{\rm f}_{7/2}$ neutron state and giving important contribution to the GTR width is restricted by the condition
$\epsilon_{\nu h''}  \approx   \epsilon_{\nu 1{\rm f}_{7/2}} + \omega_{nL} - \Delta  E $.
Since the $\nu 1{\rm f}_{7/2}$ state is isolated (cf. Fig. \ref{Ca48SP}), this relation can only be satisfied by coupling the $\nu 1{\rm f}_{7/2}$ state with itself through positive parity phonons.
%\textcolor{blue}
{In conclusion, the positive parity low-lying phonons rather than the negative parity ones give important contributions to the width because the particle and hole states of the dominant transition are isolated with other single-particle states or close to the states with the same parity, since the energy of low-lying phonons is usually similar to the energy shift $\Delta E$.}

More generally,  we can conclude from the previous discussion that when the GTR  wavefunction
is dominated  by a strong $\nu l_{j_>} \rightarrow \pi l_{j_<}$ transition, the intermediate proton particle states $\pi p''$ or neutron hole states $\nu h''$  and the phonons
will obey the following relation if the corresponding diagram gives important contributions to the GTR spreading width,
\begin{equation}
  \epsilon_{\pi p''}  \approx \epsilon_{\pi lj_<} - \omega_{nL} + \Delta E \quad
   {\rm and } \quad \epsilon_{\nu h''}  \approx \epsilon_{\nu lj_>} + \omega_{nL} -\Delta E,
\label{condition}
\end{equation}
where $\Delta E$ is the energy difference between the energy of p-h configuration and the energy of the GTR peak (approximated by the energy of RPA peak).

%---------------------------------------------------------------------------------------------------------
\begin{figure*}
\includegraphics[scale=0.3,angle=0]{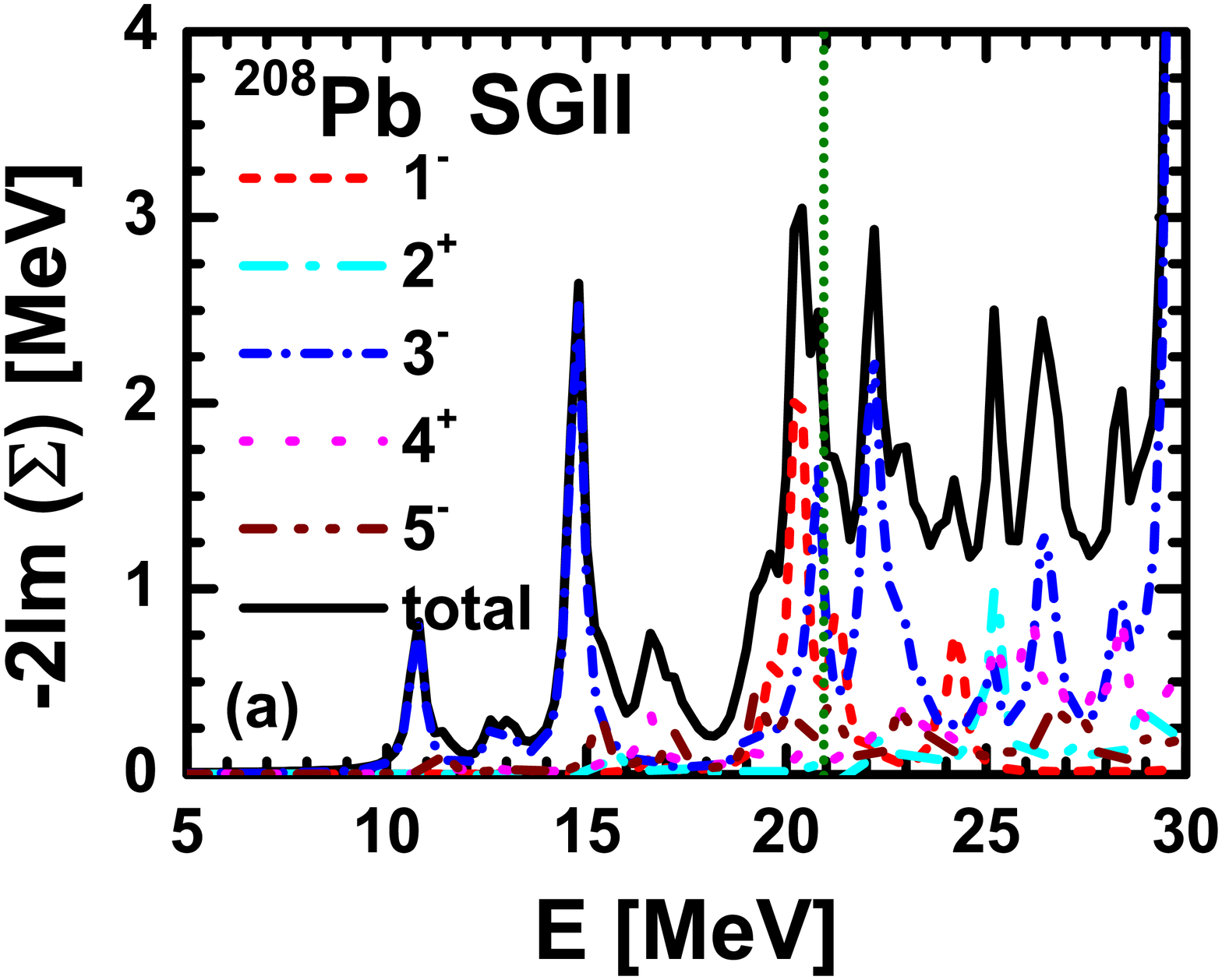}
\includegraphics[scale=0.3,angle=0]{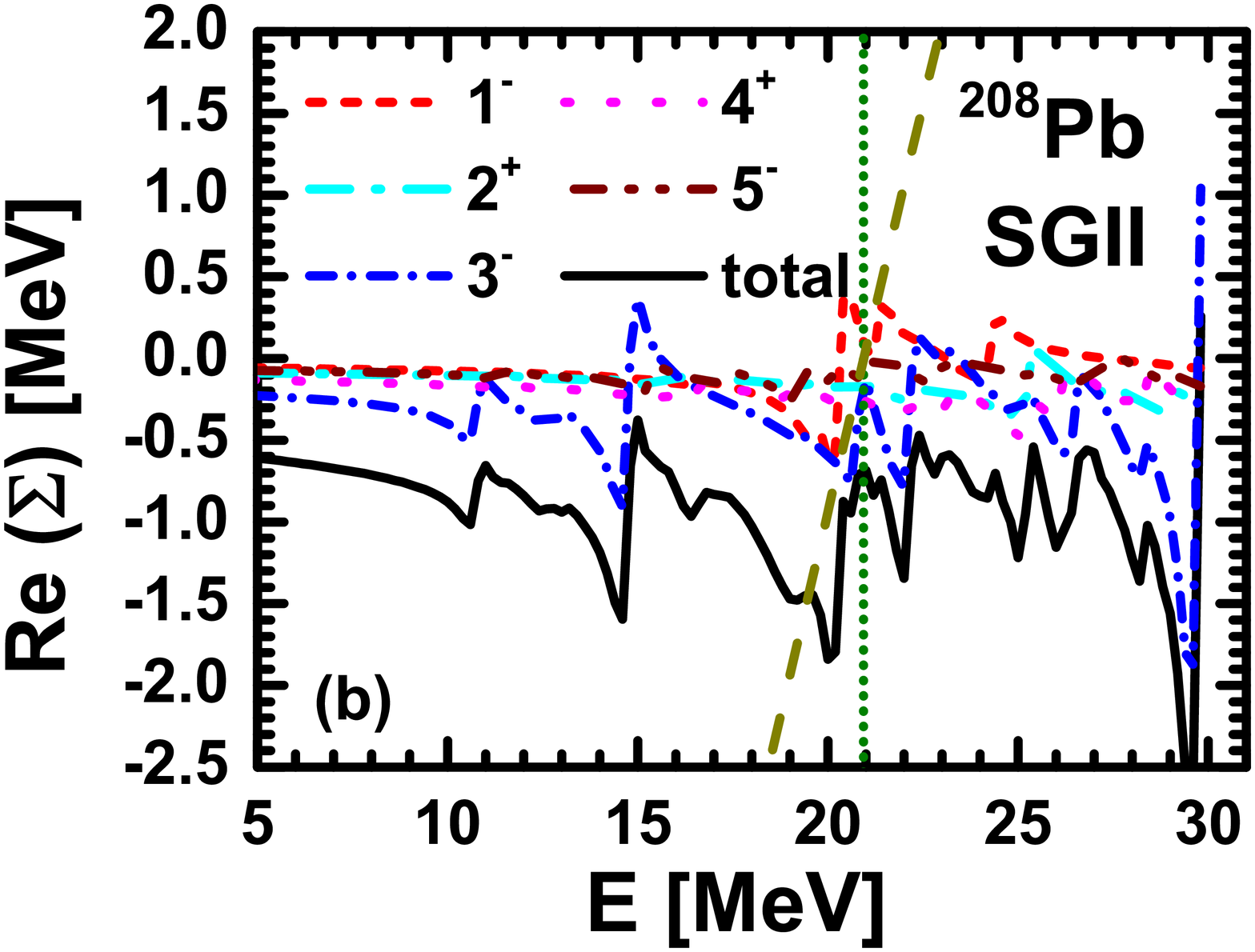}
\caption{(Color online) The same as Fig. \ref{fig7} in the case of the nucleus
$^{208}$Pb.  } \label{fig8}
\end{figure*}
%------------------------------

We now turn to the case  of $^{208}$Pb. In Fig. \ref{fig8}, we display the
values of the imaginary part and real part of the self-energy for
the GTR state from RPA calculation, in the same fashion as we did for $^{48}$Ca.
The zeros or the minima of the function
$| E -E_{\rm GTR}-{\rm Re}[\Sigma_{\rm GTR}(E)]|$, in the present case lie
at $E= 19.4$ and 20.4 MeV [cf. panel(b) of Fig. \ref{fig8}],
that correspond well
to the peak energies (19.2 and 20.6 MeV) reported in Table  \ref{table0}. From panel (a) of Fig. \ref{fig8}, we obtain that the widths at these two peak energies are 1.0 and 2.2 MeV, which again
are in agreement with the FWHMs (1.2 and 2 MeV)  reported in Table  \ref{table0}.

The individual contributions to the self-energy  from the various phonon
multipolarities  are also shown in Fig. \ref{fig8}, and the corresponding
values of the imaginary part of the self-energy calculated at the RPA energy
$E_{\rm RPA}$ are given in Table  \ref{table3}.
Phonons of positive parity give negligible contributions in this nucleus.
This is mainly due to the cancellation between
self-energy diagrams [(1) and (2)] and the phonon exchange
diagrams [(3) and (4)] of Fig. \ref{fig0}. This cancellation, which does not play so important role in $^{48}$Ca, becomes more severe for large single-particle angular momentum, which occurs in the case of $^{208}$Pb \cite{Bortignon1981}.
We also observe that the contribution from the
$1^-$ phonons to the GTR width (0.44 MeV)
is comparable to that from $3^-$ (1.05 MeV) and $5^-$ phonons (0.30 MeV).
The $1^-$ contribution is also important in $^{78}$Ni and in $^{132}$Sn.
This may appear surprising, if one sticks to the general prejudice that the largest
effects usually  originate from low-energy phonons.

\begin{table}
\caption{
Contributions to self-energy $({\cal A}_1)^{\rm GTR,GTR}_{php'h'} (E_{\rm RPA})$ arising from the two most important particle-hole configurations  forming the GTR state in $^{208}$Pb, and associated with the coupling to $1^-$ phonons. The calculation is performed
with the interaction SGII and smearing parameter $\Delta=0.2$ MeV. The self-energy is calculated at the RPA energy of the GTR. $i''$ labels the intermediate particle or hole state of the diagrams, and $E_{1^-}$ is the energy of the $1^-$ phonon state. For given $ph$ and $p^\prime h^\prime$ configurations, first we provide the total value of $W^{\downarrow}_{ph,p'h'}$ as well as the value of $({\cal A}_1)^{\rm GTR,GTR}_{php'h'} (E_{\rm RPA})$. In the following  four lines, the individual  values $W_{kph,p'h'}$ ($k=1,...,4$) are given. In the fifth line we give the contribution to  $W^\downarrow$ and  $({\cal A}_1)^{\rm GTR,GTR}$ arising from the most important diagram  contributing to the width,
associated with a single intermediate state $i''=p''$ or $i''=h''$. In the last line labeled with ``Total", the value of $({\cal A}_1)^{\rm GTR,GTR}_{php'h'} (E_{\rm RPA})$ summed over all $ph,p'h'$ particle-hole  configurations of the RPA model space is given. The notation (a,b) represents the complex number $a+ib$.}
 \scriptsize{
  \begin{tabular}{llllllll}
   \hline\hline
   $ph$ & $p^\prime h^\prime$ & $X^{\rm GTR}_{ph}$ & $X^{\rm GTR}_{p'h'}$ & $i''$ & $E_{1^-}$ (MeV)
& $W^{\downarrow}_{php'h'}$ (MeV) & $({\cal A}_1)^{\rm GTR,GTR}_{php'h'}$ (MeV) \\
   \hline
   $\pi 1{\rm h}_{9/2}- \nu 1{\rm h}_{11/2} $ & $\pi 1{\rm h}_{9/2}-
\nu 1{\rm h}_{11/2} $  & -0.51 & -0.51 & & & (-0.094, -0.31) & (-0.025, -0.083)  \\
   & & & &  & & $W_1$ (-0.019, -1.88$\times 10^{-4}$) &  \\
   & & & &  & & $W_2$ (-0.076, -0.31) &  \\
   & & & &  & & $W_3$ (0,0) &  \\
  & & & &  & & $W_4$ (0,0) &  \\
   & & & & h'' $\nu 1{\rm i}_{13/2}$ & 12.63 &$ W_2$ (-0.30,-0.095) & ( 0.080,-0.025)  \\
   & & & & h'' $\nu 1{\rm i}_{13/2}$ & 13.62 &$ W_2$ (-0.37,-0.21) & (-0.099, -0.056)  \\
   $\pi 1{\rm h}_{9/2}- \nu 1{\rm h}_{11/2} $ & $\pi 1{\rm i}_{11/2} -
\nu 1{\rm i}_{13/2} $ & -0.51 & -0.76 & & &  (0.10,-0.17) & (0.040,-0.065) $\times 2$ \\
   & & & &  & & $W_1$ (0,0) &  \\
      & & & &  & & $ W_2$ (0,0) &  \\
   & & & &  & & $W_3$ (0.11, -0.17) &  \\
   & & & &  & & $W_4$ (-8.00$\times 10^{-3}$, -9.23$\times 10^{-5}$ ) &  \\
  & & & &  & 12.63 & $W_3$ (0.25,-0.078) & (0.066,-0.021) \\
  & & & &  & 13.62 & $W_3$ (-0.15,-0.084) & (-0.039,-0.022)  \\
  $\pi 1{\rm i}_{11/2} - \nu 1{\rm i}_{13/2} $ & $\pi 1{\rm i}_{11/2} -
\nu 1{\rm i}_{13/2} $ & -0.76 & -0.76 & & & (0.089, -0.10) & (0.051, -0.060)\\
    & & & &  & & $W_1$ (0.10, -0.10) &  \\
    & & & &  & &$W_2$ (-0.012, -1.38$\times 10^{-4}$) &  \\
    & & & &  & & $W_3$ (0,0) &  \\
        & & & &  & & $ W_4$ (0,0) &  \\
            & & & & p'' $\pi 1{\rm h}_{9/2}$ & 12.63 & $W_1$ (0.21,-0.064) & (0.12, -0.037)  \\
   \hline
   Total & & & &  & &  & (0.10,-0.27)  \\
   \hline\hline
 \end{tabular}
 } \label{table1}
\end{table}

%---------------------------------------------------------------------------------------------------------
\begin{figure*}
\includegraphics[scale=0.3,angle=0]{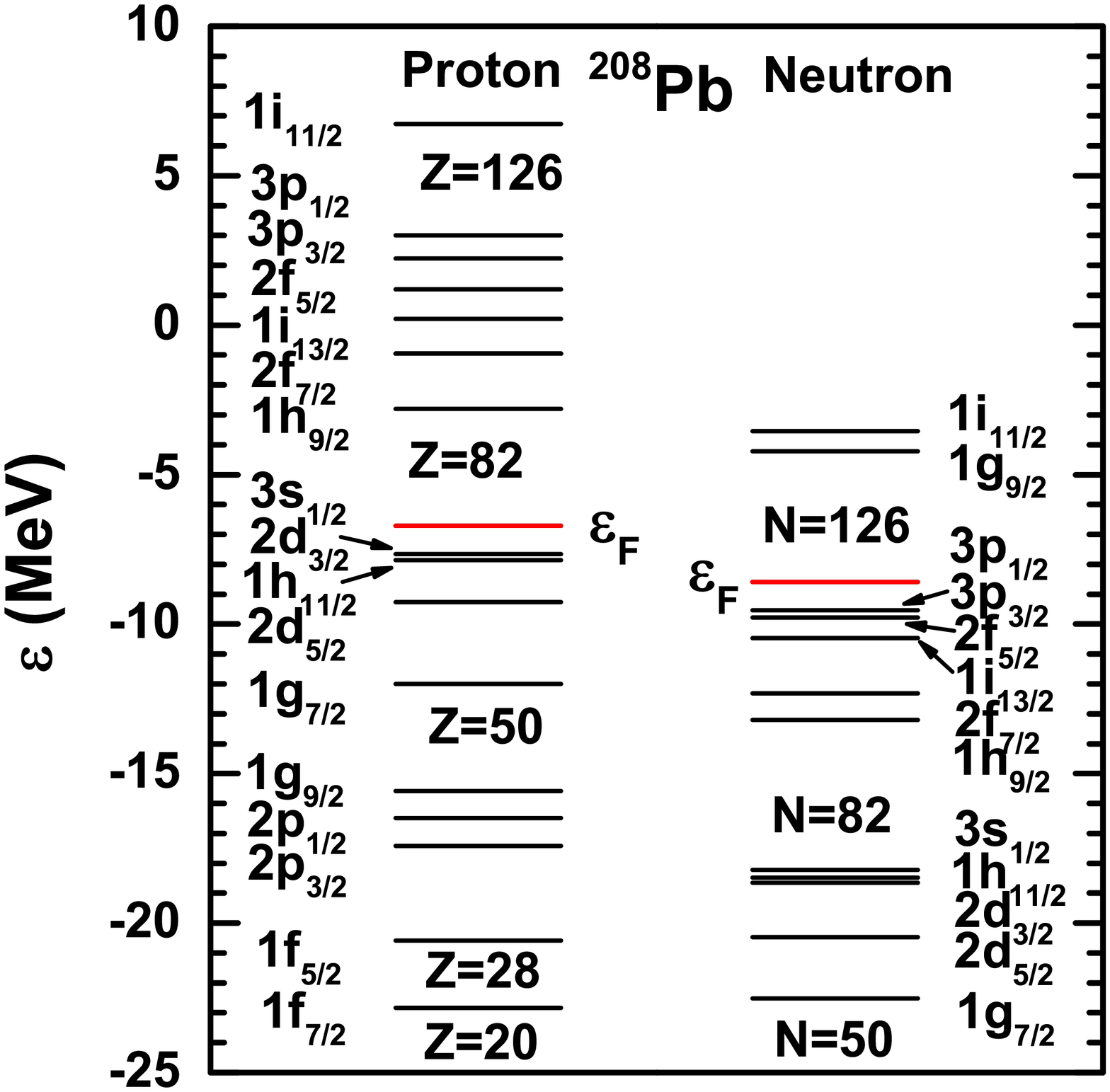}
 \caption{(Color online)  Proton and neutron single-particle spectrum in $^{208}$Pb obtained from the HF calculation with the SGII interaction.}
 \label{Pb208SP}
\end{figure*}
%---------------------------------------------------------------------------------------------------------

We study the reasons for the relevance of the $1^-$ phonons in Table \ref{table1}. There are two important p-h configurations entering the  RPA wave function of
the GTR  in $^{208}$Pb with forwards-going  RPA amplitudes in absolute value larger
than 0.5, namely $\pi 1{\rm h}_{9/2}- \nu 1{\rm h}_{11/2}$ and $\pi 1{\rm i}_{11/2} -
\nu 1{\rm i}_{13/2}$.
In Table \ref{table1} we list the contributions to the self-energy
$ ({\cal A}_1)^{\rm GTR,GTR}_{ph,p'h'} (E_{\rm RPA})$ associated with a  $ph$, $p'h'$ pair, and
arising from the coupling of these
$ph$ configurations with the  $1^-$ phonons. The leading role is played by
two collective $1^-$ phonons which carry the strongest
isovector transition strength
and have energies $E_{1^-}$= 12.63 and 13.62 MeV [i.e., isovector giant dipole resonance (IVGDR) components].
Note that due to parity conservation $W_3$ and $W_4$ vanish when $ph = p'h'$, and
$W_1,W_2$ vanish when $ph \neq p'h'$.
For the diagonal matrix element $W_{ph,ph}$
in the case of $\pi 1{\rm h}_{9/2}- \nu 1{\rm h}_{11/2}$,
the most important contribution to the width  comes from the
diagram $W_2$, associated with  the self-energy of the
hole $\nu 1{\rm h}_{11/2}$.
We can then use Eq. (\ref{condition}) with $\omega_{nL} \approx $ 13 MeV and $\Delta E=E_{\rm RPA}-(\epsilon_{\pi 1{\rm h}_{9/2}}-\epsilon_{\nu 1{\rm h}_{11/2}})=6$ MeV, and find
that the important intermediate neutron hole states coupling to the $1^-$ phonon and to the $\nu 1h_{11/2}$ hole must
lie at an  energy
$\epsilon_{\nu h''} \approx \epsilon_{\nu1h_{11/2}} +13  - 6  =  - 11$ MeV, which is
close to the energy of the ${\nu1i_{13/2}}$ hole state ($\epsilon_{\nu { i}_{13/2}}$ = -10.5 MeV).
So the intermediate hole state $\nu 1i_{13/2}$ could be coupled to $\nu 1h_{11/2}$ state and gives a non-negligible $W_2$ value. In a similar way, for the diagonal matrix element $W_{ph,ph}$ in the case of $\pi 1i_{11/2}-\nu 1i_{13/2}$, with $\Delta E= E_{\rm RPA} - (\epsilon_{\pi 1i_{11/2}}-\epsilon_{\nu 1i_{13/2}})=3.5$ MeV, we find that important
proton particle states coupling to
the $1^-$ phonon and to the $\pi 1i_{11/2}$ state should satisfy the condition
 $ \epsilon_{\pi p''}  \approx \epsilon_{\pi 1i_{11/2}} - 13  + 3.5 = - 2.5$ MeV,
 which is well fulfilled by the $\pi 1h_{9/2}$ orbital, lying at -3 MeV.
Besides these contributions associated with the diagrams $W_1$ and $W_2$, in the present case
one finds an important contribution also from diagram $W_3$
associated with the
non-diagonal matrix element of the configuration
$\pi 1{\rm i}_{11/2} - \nu 1{\rm i}_{13/2}$ and $\pi 1{\rm h}_{9/2}-
\nu 1{\rm h}_{11/2}$. The non-diagonal matrix element contributes
twice since the two different configurations could be exchanged with each other.
In conclusion, the  IVGDR phonons can give important contributions to the
GTR width when
an intermediate particle (hole) state lies below (above)
the particle (hole) state associated with an important GT
configuration by $1\hbar \omega$ energy difference, which can
be compensated by the energy of collective dipole phonon.
This condition can be satisfied in nuclei characterized by
large isospin in which the GT state is composed of two
or more important p-h configurations.

The reason for the importance of ${1^-}$ phonons  in $^{78}$Ni and $^{132}$Sn   appears to be
analogous to the case of $^{208}$Pb. In these two nuclei the
RPA wave function of the GTR  is dominated
by two p-h configurations:
$\pi 1{\rm g}_{7/2}-\nu 1{\rm g}_{9/2}$ and $\pi 1{\rm h}_{9/2}-\nu
1{\rm h}_{11/2}$
in $^{132}$Sn,
or
$\pi 1{\rm f}_{5/2}- \nu 1{\rm f}_{7/2} $ and
$ \pi 1{\rm g}_{7/2} - \nu 1{\rm g}_{9/2} $
 in $^{78}$Ni.
In the case of $^{132}$Sn  ($^{78}$Ni) the energy difference
between the GTR and the collective giant
dipole state
$E_{\rm GTR} - \omega_{nL}$ is about 1.0 MeV
(-5.3 MeV), which is close
to the energy 0.5 MeV (-6.1 MeV) of the p-h configuration
$\pi 1{\rm g}_{7/2}-\nu 1{\rm h}_{11/2}$
($\pi 1{\rm f}_{5/2}-\nu 1{\rm g}_{9/2}$),
i.e., $\epsilon_{p''}-\epsilon_{h}$ or $\epsilon_{p}-\epsilon_{h''}$.
These two nuclei are similar, also in the sense that the multipolarities $1^-$, $2^+$ and $4^+$ give comparable  contributions, and so do the $3^-$ phonons for $^{78}$Ni.
In $^{78}$Ni, the $3^-$ phonons cannot produce
spreading width through their coupling
to the most important configurations $\pi 1{\rm g}_{7/2}-\nu 1{\rm g}_{9/2}$ or $\pi 1{\rm f}_{5/2}- \nu 1{\rm f}_{7/2} $, due to the fact that the hole states
of these configurations are isolated from the other levels while the particle states are
close to the states with the same parity, as in the case of $^{48}$Ca. However, the $3^-$
phonons produce some width by coupling to the p-h configuration
$\pi 2{\rm p}_{3/2}-\nu 2{\rm p}_{3/2} $.

%-------------------------------------------------------------------------------

%---------------------------------------------------------
\section{Conclusion}\label{conclu}
%---------------------------------------------------------
Many studies of the GTR are performed at the mean-field level.
However, experiment shows that such a resonance has a conspicuous
width, coming mainly from coupling to complex nuclear configurations.
Benchmarking nuclear models through their capability to reproduce at the
same time not only energies and strengths, but also widths of the GT states, is
not very much pursued - the only exception being probably
the recent work of Ref. \cite{Litvinova2014}. In the present work,
we wish to test systematically a microscopic model, in which on top
of HF+RPA the particle-vibration coupling is introduced based
consistently on the use of a Skyrme-type force.

In this paper we have applied our model to the cases of $^{48}$Ca, $^{78}$Ni, $^{132}$Sn,  and $^{208}$Pb,
which can be well described  as doubly closed shell  nuclei.
In the future we plan to  include pairing correlations for
systematic calculations in open shell nuclei. Our results account
well for the experimental findings in $^{208}$Pb, especially
concerning the lineshape
of the GT strength. For $^{48}$Ca, the experimental width and
fragmentation is partly reproduced by the coupling with phonons.
We have made predictions for the exotic
nuclei $^{78}$Ni and $^{132}$Sn. Large
spreading widths and strong fragmentation are obtained for these two nuclei.

For $^{208}$Pb the
experimental strength integrated up to E = 25 MeV is 71\% of the
RPA+PVC result, while for $^{48}$Ca this value up to 20
MeV is 63\%.
So, we can conclude that
the coupling with phonons can produce some quenching of the main GTR,
but also other effects, like the inclusion of tensor
force and the coupling with high-energy, uncorrelated 2p-2h
configurations, need to
be considered.

The mechanism for the spreading width and fragmentation is analyzed in detail,
particularly in the case of
the two nuclei $^{48}$Ca and $^{208}$Pb.
To a large extent,
the diagonal approximation holds well in the sense that
the real part and
imaginary part of the self-energy associated with the RPA resonance
state calculated at the GTR peak energy account quite well for the
energy shift and width of GTR, respectively, produced by the particle-vibration
coupling in the full diagonalization. The
importance of phonons with different multipolarities is also discussed in
detail. The energies of phonons are important for minimizing the energy
denominators in the self-energy, whereas the reduced transition probabilities
of the phonons influence the matrix elements of the particle-vibration coupling
vertex. General arguments may suggest that low-lying phonons are
the most effective, in this respect, in particular in order
to produce small energy denominators. We have also found,
nonetheless, that in
nuclei characterized by a large neutron excess, such as $^{78}$Ni,
$^{132}$Sn, and $^{208}$Pb, the isovector giant dipole phonons can give important
contributions to the width. In fact, their energy can match
the energy difference between either the particles or holes associated with two important GT configurations.

Comparing the phonon energy and reduced transition probability between the
experiment and theory, the phonon properties of $^{208}$Pb
are best described, while in $^{48}$Ca the reduced transition probabilities
are not reproduced well. This may indicate that the difference
in the quality of the results between $^{208}$Pb on the one
side, and $^{48}$Ca on the other side,  is not due to  a breakdown of our overall
physical picture, but rather to the  inaccurate reproduction of the experimental properties of the low-lying
phonons with the adopted interaction.

%=================================================================
{\center{\bf ACKNOWLEDGMENTS}}

This work was partly supported by the National Natural Science Foundation of China
under Grants No. 11305161.

%=================================================================

%
%------------------------------------------------------------------

\clearpage
\bibliographystyle{apsrev}

%\bibliography{RPAPVC} % Produces the bibliography via BibTeX.

\end{CJK*}
\end{document}